\documentclass[12pt,a4paper]{iopart}
\usepackage{iopams}
\usepackage{setstack,cite}
\usepackage{amssymb,graphicx}

\newfont{\logo}{logo10}

\newcommand{\bea}{\begin{eqnarray}}
\newcommand{\eea}{\end{eqnarray}}
\newcommand{\ds}{\displaystyle}

\begin{document}
\bibliographystyle{revtex}

\title{Multicomponent coherently coupled and incoherently coupled solitons and their collisions}
\author{T Kanna and K Sakkaravarthi}
\address{Post Graduate and Research Department of Physics, Bishop Heber College, Tiruchirapalli - 620 017, India.}
\ead{kanna\_phy@bhc.edu.in(corresponding author)}
\ead{ksakkaravarthi@gmail.com}

\date{\today}
\begin{abstract}
We consider the integrable multicomponent coherently coupled nonlinear Schr\"odinger (CCNLS) equations describing simultaneous propagation of multiple fields in Kerr type nonlinear media. The correct bilinear equations of $m$-CCNLS equations are obtained by using a non-standard type of Hirota's bilinearization method and the more general bright one solitons with single hump and double hump profiles including special flat-top profiles are obtained. The solitons are classified as coherently coupled solitons and incoherently coupled solitons depending upon the presence and absence of coherent nonlinearity arising due to the existence of the co-propagating modes/components. Further, the more general two-soliton solutions are obtained by using this non-standard bilinearization approach and various fascinating collision dynamics are pointed out. Particularly, we demonstrate that the collision among coherently coupled soliton and incoherently coupled soliton displays a non-trivial collision behaviour in which the former always undergoes energy switching accompanied by an amplitude dependent phase-shift and change in the relative separation distance, leaving the latter unaltered. But the collision between coherently coupled solitons alone is found to be standard elastic collision. Our study also reveals the important fact that the collision between incoherently coupled solitons arising in the $m$-CCNLS system with $m=2$ is always elastic, whereas for $m>2$ the collision becomes intricate and for this case the $m$-CCNLS system exhibits interesting energy sharing collision of solitons characterized by intensity redistribution, amplitude dependent phase-shift and change in  relative separation distance which is similar to that of the multicomponent Manakov soliton collisions. This suggests that the $m$-CCNLS system can also be a suitable candidate for soliton collision based optical computing in addition to the Manakov system.
\end{abstract}

\vspace{1.5cm}

Journal Reference: {\emph{J. Phys. A: Math. Theor.} \textbf{44} (2011) 285211}

\maketitle

\section{Introduction}
Multicomponent solitons/solitary waves have attracted considerable attention in the field of nonlinear science as they display a rich variety of propagation and collision properties which are not possible in their single component counterparts \cite{{kiv},{akmbook},{molle},{tkprl},{tkpre},{tkpre2},{tkpra},{tkepj},{tkjpa},{man},{rk},{sov},{park},{wadati},{yang1},{abloinv},{akmprl},{degas}}. Such solitons appear in different areas of science like nonlinear optics \cite{{kiv},{molle},{akmbook}}, Bose-Einstein condensates \cite{bec}, bio-physics \cite{scot}, plasma physics \cite{akmbook}, etc. Here, our main focus is on nonlinear optics. In this context, multicomponent temporal solitons can be formed when an optical pulse propagating through a multimode fiber due to a delicate balance between dispersion and Kerr nonlinearity \cite{kiv}. Multicomponent spatial solitons are self-trapped optical beams that result from an interplay between diffraction and nonlinearity \cite{kiv}.

Mathematically, the propagation and collision properties of multicomponent solitons/solitary waves arising in the field of nonlinear optics can be well described within the framework of multicomponent nonlinear Schr\"odinger (NLS) type equations \cite{{kiv},{molle}}. Especially, the short pulse propagation in polarization maintaining multimode birefringent fiber is governed by a set of multicomponent incoherently coupled NLS (ICNLS) equations \cite{crosi}. Similar set of ICNLS equations also arises in the context of partially incoherent beam propagation in Kerr type nonlinear media \cite{akmprl}. These ICNLS equations involve the nonlinear couplings due to self-phase modulation (SPM) and cross-phase modulation (XPM) and depend only on the local intensities of the co-propagating fields, but insensitive to their phases\cite{kiv}.

In general cases, like pico-second pulse propagation in non-ideal low birefringent multimode fibers or beam propagation in weakly anisotropic Kerr type nonlinear media, the coherent effects due to the interaction of co-propagating fields should also be considered \cite{{kiv},{crosi}}. To be specific, the propagation of coherently coupled orthogonally polarized waveguide modes in Kerr type nonlinear medium is governed by the following $2$-component coherently coupled NLS (CCNLS) type equations \cite{kiv}:
\bea
i q_{1,z}+\delta q_{1,tt}-\mu q_{1}+ (|q_{1}|^2+\sigma |q_{2}|^2)q_{1}+\lambda q_2^2 q_{1}^*=0,\nonumber\\
i q_{2,z}+\delta q_{2,tt}+\mu q_{2}+ (\sigma |q_{1}|^2+|q_{2}|^2)q_2+\lambda q_{1}^2 q_{2}^*=0,
\label{e1}
\eea
where $q_1$ and $q_2$ are slowly varying complex amplitudes in each polarization mode, $z$ and $t$ are the propagation direction and transverse direction, respectively, $\mu$ is the degree of birefringence. Here, the incoherent and coherent couplings are represented by the parameters $\sigma$ and $\lambda$, respectively. The above equation (\ref{e1}) also arises in the context of beam propagation in isotropic Kerr type nonlinear gyrotropic medium \cite{gyro}. In  equation (\ref{e1}), the terms $q_2^2 q_1^*$ and $q_1^2 q_2^*$ correspond to four wave mixing (FWM) process which arise due to the coherent coupling between the co-propagating fields.

Generally, these CCNLS equations and also the ICNLS equations are non-integrable. However, these become integrable for specific choices which are of physical significance \cite{{zak},{park},{saha},{zhang}}. In recent years, much attention has been paid to the integrable and non-integrable coupled ICNLS equations and many interesting phenomena have been explored \cite{kiv}. To be specific, the integrable $m$-component Manakov type equations, with arbitrary $m$, are well studied and it has been pointed out that these equations support bright optical solitons which undergo fascinating energy sharing collisions that have immediate technological applications in the context of collision based optical computing \cite{{tkprl},{rk},{sov},{ste1},{ste2}} and also in soliton amplification \cite{tkpre2}. Very recently, CCNLS equations have also been started to receive renewed attention due to their rich structure \cite{{park},{zhang},{tkjpa},{chow}}. Particularly, a set of physically interesting integrable $2$-component CCNLS equations related to \eref{e1} is
\bea
i q_{1z}+ q_{1tt}+\gamma (|q_1|^2+2 |q_2|^2)q_1-{\gamma} q_2^2 q_1^*=0,\nonumber\\
i q_{2z}+ q_{2tt}+\gamma (2|q_1|^2+ |q_2|^2)q_2-\gamma q_1^2 q_2^*=0,
\label{e2}
\eea
where $\gamma$ is the coupling coefficient. In gyrotropic nonlinear medium, the above equation (2) can be obtained for the following choice of the susceptibility tensor $\chi^{(3)}$, with its components satisfying the relation $\chi_{xxxx}^{(3)}=\chi_{xxyy}^{(3)}=\chi_{xyxy}^{(3)}=\chi_{xyyx}^{(3)}$ \cite{gyro}. Equation (2) also describes the propagation of two optical pulses in an isotropic nonlinear Kerr medium when the components $\chi_{xxyy}^{(3)}$, $\chi_{xyxy}^{(3)}$ and $\chi_{xyyx}^{(3)}$ of the susceptibility tensor $\chi^{(3)}$ can be expressed as  $\chi_{xxyy}^{(3)}+\chi_{xyxy}^{(3)}=-2\chi_{xyyx}^{(3)}$ \cite{park}.

Apart from the 2-component CCNLS equations, $m$-component CCNLS equations with $m>2$ are also of special physical interest and have been derived under different physical contexts. Particularly, the spatial evolution of mutually guided four wave mixing states in $\chi^{(3)}$ medium is governed by $3$-component CCNLS type equation \cite{ander}. It has also been shown that the co-propagation of two optical pulses in birefringent fiber can be described by 4-component CCNLS equations \cite{4cnls}. In ref. \cite{wadati}, the dynamics of spinor Bose-Einstein condensates has been investigated by considering a set of integrable 3-coupled CCNLS type equations \cite{tkjmp} and novel polar and ferromagnetic solitons have been reported. Now it is of interest to investigate the integrable multicomponent CCNLS equations which are closely associated with the near-integrable or non-integrable systems appearing in nonlinear optics.

Being motivated by these reasons, we consider the following integrable $m$-component generalization of (\ref{e2}) describing the simultaneous propagation of $m$-optical fields in Kerr type nonlinear media.
\bea
\hspace{-2cm} i q_{j,z}+ q_{j,tt}+\gamma \left(|q_j|^2+2 \sum_{l=1,l\neq j}^m |q_l|^2\right) q_j - {\gamma} \sum_{l=1,l\neq j}^m q_l^2 q_j^*=0, \quad ~~j=1,2,3,...,m.
\label{e2nc}
\eea
The above system has been studied in ref.\cite{zhang} by applying the Hirota's direct method but trivial soliton solutions with less number of parameters only have been reported due to the restricted bilinearization of (\ref{e2nc}). It also should be noticed that the information regarding the coherent and incoherent contributions from the co-propagating fields are lost completely if the two-soliton solution is constructed by a linear superposition as pointed out in refs.\cite{{park},{zhang}}. So it is of importance to obtain correct bilinear equations of system (\ref{e2nc}) which will result in more general soliton solutions with interesting properties. In ref. \cite{tkjpa}, Kanna \etal have considered a 2-component integrable model which can be reduced from (\ref{e2}) by redefining $q_1$ as $i q_1$ and reported novel solitons with variable profiles and classify them as degenerate (solitons with same intensity in both components) and non-degenerate (solitons with different intensity in two components) solitons. But study on the present system (3) suggests that a broader classification of solitons of the general $m$-component system (\ref{e2nc}), with arbitrary $m$, can be made based on the presence and absence of the coherent nonlinearity, and the degenerate and non-degenerate solitons reported in \cite{tkjpa} appear as their sub-cases, which will be discussed in the following sections.

The aim of the present work is three-folded. First, to obtain the correct bilinear equations of system (\ref{e2nc}) and to construct exact one- and two-bright soliton solutions of (\ref{e2nc}). Next, to analyse the collision dynamics of solitons in the 2-component and 3-component CCNLS equations and to bring out their salient features. Finally, to generalize the results to the $m$-component case.

This paper is set out as follows. The correct bilinear equations of (\ref{e2nc}) are obtained and the solitons are classified in a systematic way in section 2. The bright one- and two-soliton solutions of $2$-component and $3$-component CCNLS equations are obtained in section 3 and in section 4, respectively. Then the results are generalized to arbitrary $m$-component case in section 5. Section 6 deals with the collision dynamics of the solitons. Final section is allotted for conclusion.

\section{Non-standard bilinearization and classification of solitons of integrable multicomponent CCNLS system (\ref{e2nc})}
\indent Hirota's direct bilinearization method is one of the powerful techniques to construct soliton solutions of integrable nonlinear evolution equations \cite{hir}. In this section, we construct the bilinear equations of the CCNLS system (\ref{e2nc}) by applying the Hirota's direct method \cite{hir}. A new type of bilinearization procedure has been developed by introducing an auxiliary function for the Sasa-Satsuma higher order nonlinear Schr\"odinger equations in ref. \cite{gil} by Gilson \emph{et al}. By adopting this technique, here we obtain correct bilinear equations of system (\ref{e2nc}) resulting in more general bright soliton solutions which display the effects of both intensity and phase dependent nonlinearities. By performing the bilinearizing transformation
\bea
q_j=\frac{g^{(j)}}{f},\quad j=1,2,...,m,
\eea
to equation (\ref{e2nc}) with the introduction of an auxiliary function $s$, we arrive at the following set of bilinear equations:
\numparts\label{be}\bea
D_1(g^{(j)}\cdot f) &=& \gamma s g^{(j)*}, \quad\quad\quad\quad j=1,2,...,m,\\
D_2(f \cdot f) &=& 2 \gamma \left(\sum_{j=1}^m |g^{(j)}|^2\right), \\
s\cdot f&=&\sum_{j=1}^m (g^{(j)})^2, \label{3beq}
\eea
\endnumparts
where $D_1=iD_z+D_t^2$ and $D_2=D_t^2$. Here $g^{(j)}$ and $f$ are complex and real functions, respectively, $*$ denotes the complex conjugate, $D_z$ and $D_t$ are the well known Hirota's $D$-operators \cite{hir} which are defined as
\bea
D_z^{p}D_t^{q}(a\cdot b) =\bigg(\frac{\partial}{\partial z}-\frac{\partial}{\partial z'}\bigg)^p\bigg(\frac{\partial}{\partial t}-\frac{\partial}{\partial t'}\bigg)^q a(z,t)b(z',t')\Big|_{ \ds (z=z', t=t')}.\nonumber
\eea
The above set of equations (5) can be solved by introducing the following power series expansions for $g^{(j)}$, $f$, and $s$
\numparts\bea
&&g=\chi g^{(j)}_1+\chi^3 g^{(j)}_3+\ldots,  \quad\quad j=1,2,...,m,\\
&&f=1+\chi^2 f_2+\chi^4 f_4+\ldots,\quad  s=\chi^2 s_2+\chi^4 s_4+\ldots,
\eea\endnumparts\label{ps}
where $\chi$ is the formal power series expansion parameter.  The resulting set of linear partial differential equations after collecting the terms with the same powers in $\chi$, can be solved recursively to obtain the forms of $g^{(j)}$, $f$, and $s$.

It can be inferred from the above  bilinear equations (5) that when the auxiliary function ``$s$" becomes zero, the contribution from the coherent coupling vanishes and the above bilinear equations reduce to that of integrable $m$-component Manakov system. We notice that for $s=0$ equation (\ref{3beq}) results in the condition $\displaystyle\sum_{j=1}^m (g^{(j)})^2=0$, which ultimately restricts the energy sharing of a given field/soliton among all its components. In the following, we obtain explicit conditions on soliton parameters for which ``$s$" becomes zero and we refer the soliton arising for this choice, $s=0$, as incoherently coupled soliton (ICS) as the contribution from the coherent nonlinearity is absent. The ICS results due to the interplay between dispersion/diffraction and the nonlinearity arising due to SPM and XPM effects. However the general  bilinear equations (5) with non-vanishing auxiliary function ``$s$"involve the effect of coherent coupling also. Hence, we designate the soliton resulting for the general choice, $s\neq0$, as coherently coupled soliton (CCS). These CCSs are formed due to the contribution from the dispersion/diffraction and the combined nonlinear effect resulting from SPM, XPM and four wave mixing process.

\section{Bright soliton solutions of 2--component CCNLS equations}
In this section, the bright one- and two-soliton solutions of the 2-component CCNLS equations (\ref{e2}) are obtained by applying the non-standard type of Hirota's bilinearization method explained in the previous section. We present the results for the $m=2$ and $m=3$ cases explicitly in order to emphasize the additional features of the $m= 3$ case ($m$ represents the components). Here onwards we designate the $m$-component $N$-soliton solution as ($m,N$) soliton solution for convenience.

\subsection{\textbf{Bright (2,1) soliton solution}}
\begin{figure}[h]
\centering\includegraphics[width=0.65\columnwidth]{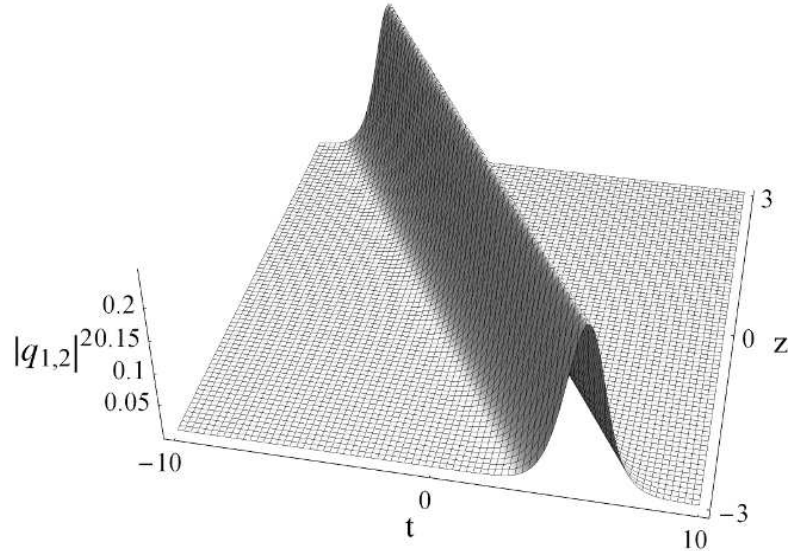}
\caption{Degenerate (2,1) incoherently coupled soliton.}
\label{2cos1}
\end{figure}

\begin{figure}[h]
\centering\includegraphics[width=0.65\linewidth]{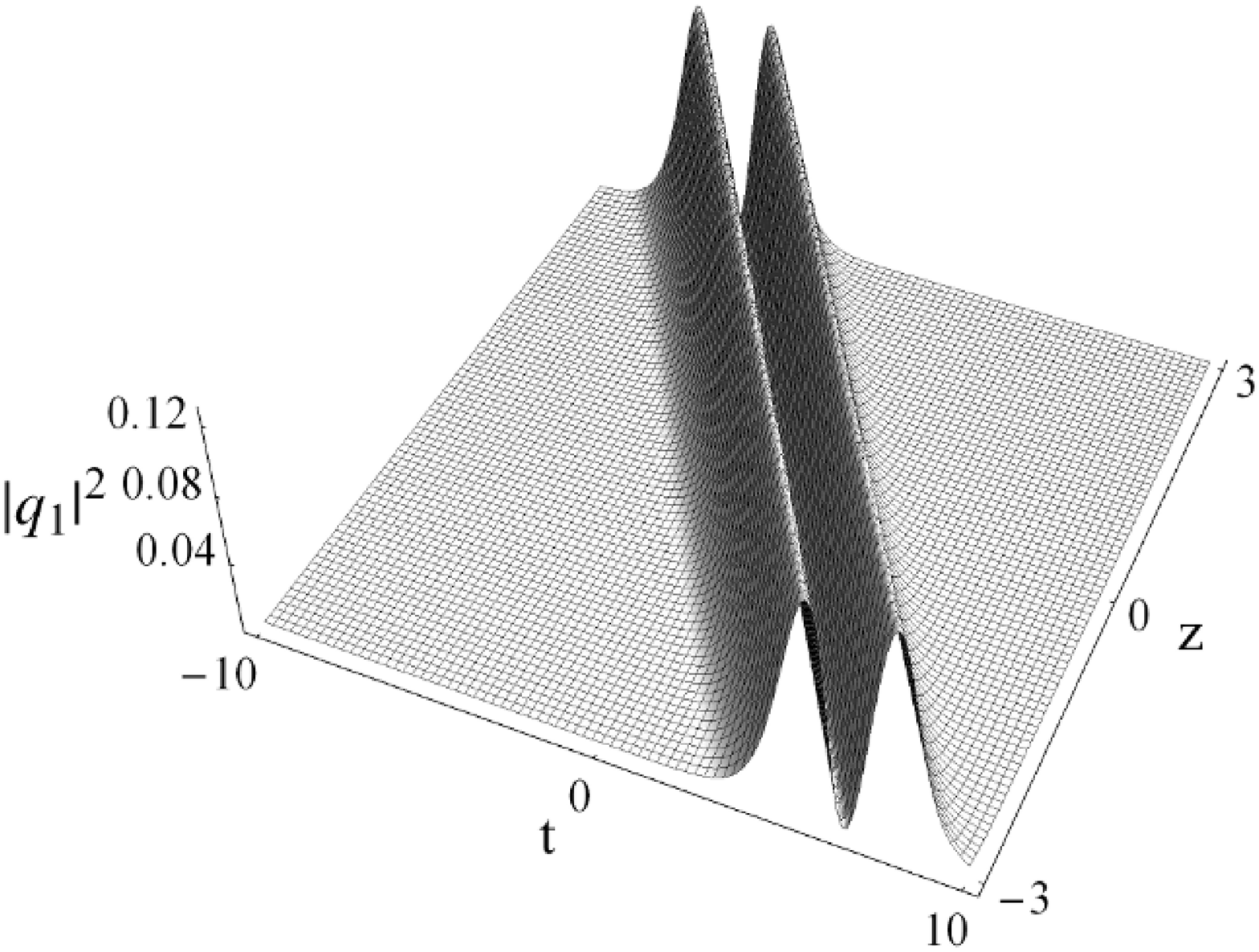}\\\includegraphics[width=0.65\linewidth]{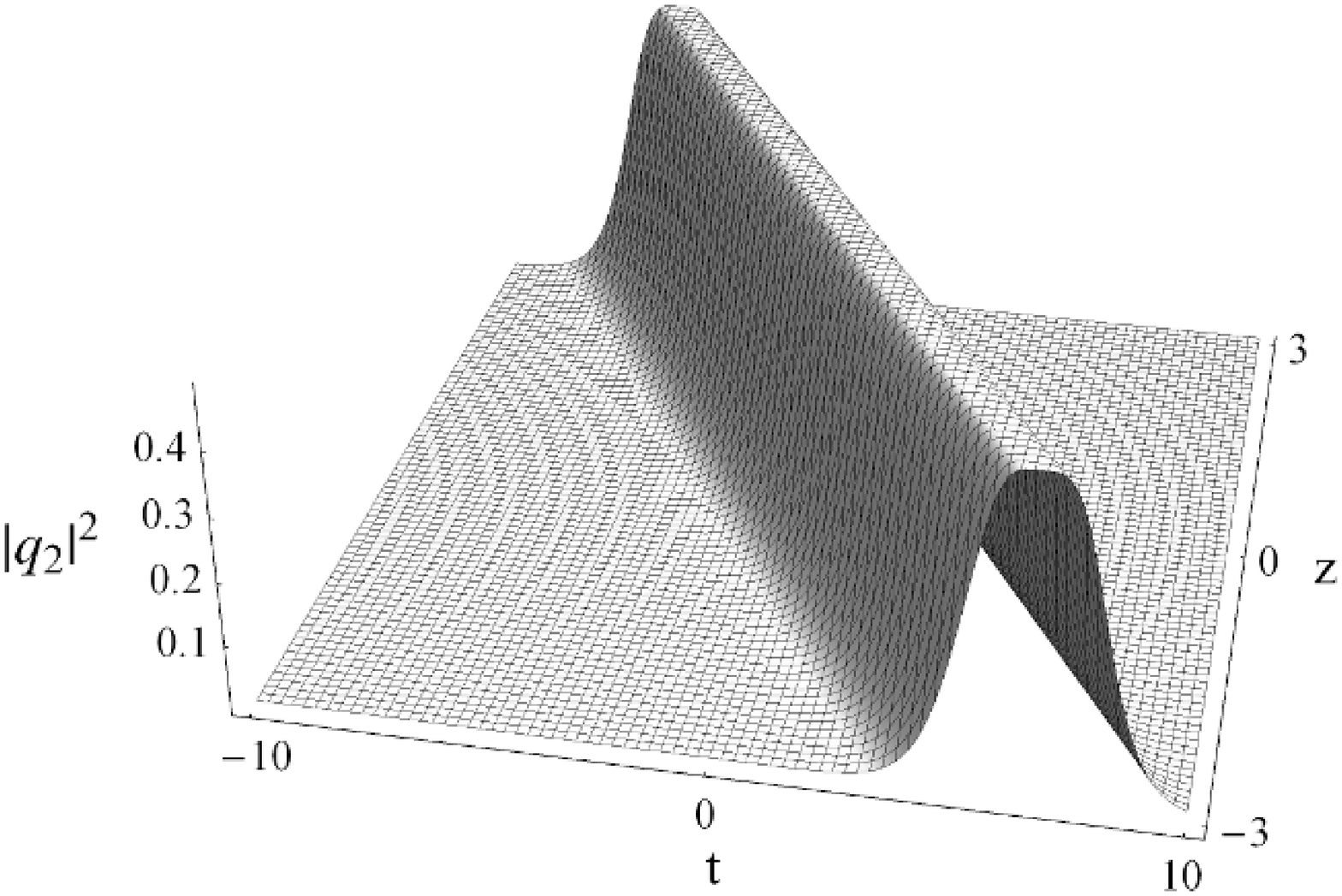}
\caption{Non-degenerate ($2,1$) coherently coupled soliton with double hump and flat-top profiles.}
\label{2cos2}
\end{figure}

To obtain bright one-soliton solution of system (2), we restrict the power series expansion (6) as $g^{(j)}=\chi g_1^{(j)}+\chi^3 g_3^{(j)}$, $j=1,2$, $f=1+\chi^2 f_2+\chi^4 f_4$, $s=\chi^2 s_2$. By substituting these series expansions into (5) and after recursively solving the equations resulting at like powers of $\chi$, we obtain the following one-soliton solution.
\numparts
\bea
q_j&=&\frac{\alpha_1^{(j)} e^{\eta_1}+e^{2\eta_1+\eta_1^*+\delta_{11}^{(j)}}}{1+e^{\eta_1+\eta_1^*+R_1}+e^{2\eta_1+2\eta_1^*+\epsilon_{11}}},\quad j=1,2,
\eea

\noindent where
\bea
&&e^{\delta_{11}^{(j)}}=\frac{\gamma \alpha_1^{(j)*} ((\alpha_1^{(1)})^2+(\alpha_1^{(2)})^2)}{2 (k_1+k_1^*)^2},\quad
e^{R_1}=\frac{\kappa_{11}}{(k_1+k_1^*)}, \quad j=1,2,\\
&&~e^{\epsilon_{11}}=\frac{\gamma^2 \big|(\alpha_1^{(1)})^2+(\alpha_1^{(2)})^2\big|^2}{4 (k_1+k_1^*)^4}, \quad\quad \kappa_{11}=\frac{\gamma {(|\alpha_1^{(1)}|^2+|\alpha_1^{(2)}|^2)}}{(k_1+k_1^*)},
\eea
The auxiliary function $s$ is found to be
\bea
s&=&((\alpha_1^{(1)})^2+(\alpha_1^{(2)})^2) e^{2\eta_1}. \label{s1s}
\eea \label{1s}\endnumparts
Here, $\eta_1=k_1(t+i k_1 z)$, $k_1=k_{1R}+i k_{1I}$, and $\alpha_1^{(j)}$'s are complex parameters. Throughout this paper, the real and imaginary parts of a parameter are represented by the subscripts R and I, respectively.
\\
\noindent (i) \underline{Bright (2,1) ICS:} \\
\indent The (2,1) ICS results for the vanishing auxiliary function ($s=0$). We find from (\ref{s1s}) that the condition for $s$ to be zero is $(\alpha_1^{(1)})^2+(\alpha_1^{(2)})^2=0$. This (2,1) ICS always exhibits the standard ``sech" type profile and can be expressed as
\bea
q_j&=& A_j~\mbox{sech}\left(\eta_{1R}+R_1/2\right)e^{i\eta_{1I}}, \quad j=1,2,
\label{ms}
\eea
where $A_j=\frac{\alpha_1^{(j)} k_{1R}}{\sqrt{2\gamma |\alpha_1^{(1)}|^2}}$, $R_1=\mbox{ln}\left(\frac{\kappa_{11}}{2 k_{1R}}\right)$, $\eta_{1R}=k_{1R}(t-2k_{1I}z)$, and $\eta_{1I}=k_{1I}t+(k_{1R}^2-k_{1I}^2)z$. For this case, either $\alpha_1^{(2)}=\pm i\alpha_1^{(1)}$ or $\alpha_1^{(1)}=\pm i\alpha_1^{(2)}$ and correspondingly the solitons in $q_1$ and $q_2$ components are related as $q_2=\pm iq_1$ or $q_1=\pm iq_2$. Ultimately, the intensity profiles of these ICSs are same in both the components (that is, $|q_1|^2=|q_2|^2$). One can also refer these equal intensity solitons in both components as degenerate (2,1) ICSs and are characterized by two complex parameters $k_1$ and $\alpha_1^{(1)}$ (or $\alpha_1^{(2)}$). These solitons behave as standard NLS solitons during propagation. The amplitude of soliton in the $q_j$-th component is $A_j$. The velocity and central position of soliton in both components are $2k_{1I}$ and $\frac{R_1}{2k_{1R}}$, respectively, and can be tuned by altering either $k_1$ or $\alpha_1^{(1)}$ (or $\alpha_1^{(2)}$). Such an incoherently coupled soliton is depicted in figure \ref{2cos1} for the parameters $\gamma=2$, $k_1=1-i$, $\alpha_1^{(1)}=1+i$, and $\alpha_1^{(2)}=1-i$.
\\
\noindent (ii) \underline{Bright (2,1) CCS:} \\
\indent The ($2,1$) CCS solution can be obtained for non-zero auxiliary function ($s\neq0$). We require $(\alpha_1^{(1)})^2+(\alpha_1^{(2)})^2\neq0$, for non-vanishing $s$ and the corresponding ($2,1$) CCS solution is found to be
\bea
q_j=2A_j\left(\frac{\mbox{cos}(P_j)~\mbox{cosh}\left(Q\right)
+i~\mbox{sin}(P_j)~\mbox{sinh}(Q)}{4 \mbox{cosh}^2(Q)+L}\right)e^{i\eta_{1I}}, \quad j=1,2,
\label{os}
\eea
where $A_j=e^{{\frac{l_j+\delta_{11}^{(j)}-\epsilon_{11}}{2}}}$, $P_j=\frac{\delta_{11I}^{(j)}-l_{jI}}{2}$, $l_{j}=\ln({\alpha_1^{(j)}})$, $j=1, 2$, $Q=\eta_{1R}+\frac{\epsilon_{11}}{4}$, $L=e^{({R_1-\frac{\epsilon_{11}}{2}})}-2$, $\eta_{1R}=k_{1R}(t-2k_{1I}z)$, and $\eta_{1I}=k_{1I}t+(k_{1R}^2-k_{1I}^2)z$. The (2,1) CCS can exhibit both equal and non-equal intensities in both components, which may be referred as degenerate and non-degenerate (2,1) CCSs, respectively. Thus the degenerate and non-degenerate solitons obtained by one of the authors and co-workers in ref. \cite{tkjpa} can be deduced as sub-cases of coherently coupled solitons of system (\ref{e2}) discussed here. Generally, these CCSs admit double hump profiles. The existence of coherent coupling is reflected by such kind of distinct profiles. One can also obtain perfect ``sech" type soliton profile when the parameters are chosen suitably. This can be achieved for the condition $\alpha_1^{(1)*}\alpha_1^{(2)} - \alpha_1^{(1)}\alpha_1^{(2)*}=0$,  which makes $P_j=L=0,~j=1,2$. Here $A_j$ represents the amplitude (peak value of the envelope) of soliton in $q_j$-th component and $2k_{1I}$ and $\frac{\epsilon_{11}}{4k_{1R}}$ are the velocity and the central position of the soliton in both components, respectively. These (2,1) CCSs are characterized by three complex parameters $k_1$, $\alpha_1^{(1)}$ and $\alpha_1^{(2)}$. The non-degenerate (2,1) CCS having double hump profile in $q_1$ component and special flat-top profile in $q_2$ component is depicted in figure \ref{2cos2} for the parameters $\gamma=2$, $k_1=1-i$, $\alpha_1^{(1)}=1$, and $\alpha_1^{(2)}=1.4i$. Such flat-top type solitons have been reported in non-integrable complex Ginzburg-Landau equations \cite{akmbook}. Thus from our above analysis we observe that one can switch from coherently coupled soliton to incoherently coupled soliton and vice-versa  by tuning the polarization parameters ($\alpha^{(j)}$'s) suitably.

\subsection{\textbf{Bright (2,2) soliton solution}}
The bright two-soliton solution of system (\ref{e2}) is obtained by restricting the power series expansion (6) as $g^{(j)}=\chi g_1^{(j)}+\chi^3 g_3^{(j)}+\chi^5 g_5^{(j)}+\chi^7 g_7^{(j)}$, $j=1,2$, $f=1+\chi^2 f_2+\chi^4 f_4+\chi^6 f_6+\chi^8 f_8$, $s=\chi^2 s_2+\chi^4 s_4+\chi^6 s_6$. Then, by solving the resultant set of linear partial differential equations, we get the bright two-soliton solution as
\numparts\bea
q_j=\frac{g^{(j)}}{f},\quad j=1,2,
\eea
where
\bea
\hspace{-2.5cm}g^{(j)}=&&\alpha _1^{(j)} e^{\eta _1} +\alpha _2^{(j)} e^{\eta _2}+e^{2 \eta _1+\eta _1^{*}+\delta_{11}^{(j)}}+e^{2 \eta _1+\eta _2^{*}+\delta_{12}^{(j)}}
+e^{2 \eta _2+\eta _1^{*}+\delta_{21}^{(j)}}+e^{2 \eta _2+\eta _2^{*}+\delta_{22}^{(j)}}\nonumber\\
\hspace{-2.5cm}&&+e^{\eta_1+\eta_1^*+\eta_2+\delta_1^{(j)}}+e^{\eta_2+\eta_2^*+\eta_1+\delta_2^{(j)}}+e^{2\eta_1+2\eta_1^*+\eta_2+\mu_{11}^{(j)}}+e^{2\eta_1+2\eta_2^*+\eta_2+\mu_{12}^{(j)}}\nonumber\\
\hspace{-2.5cm}&&+e^{2\eta_2+2\eta_1^*+\eta_1+\mu_{21}^{(j)}}+e^{2\eta_2+2\eta_2^*+\eta_1+\mu_{22}^{(j)}}+e^{2\eta_1+\eta_1^*+\eta_2+\eta_2^*+\mu_1^{(j)}}\nonumber\\
\hspace{-2.5cm}&&+e^{2\eta_2+\eta_2^*+\eta_1+\eta_1^*+\mu_2^{(j)}}+e^{2\eta_1+2\eta_1^*+2\eta_2+\eta_2^*+\phi_1^{(j)}}+e^{2\eta_1+2\eta_2+2\eta_2^*+\eta_1^*+\phi_2^{(j)}},\quad j=1,2,\\
\hspace{-2.5cm}f=&&1+e^{\eta_1+\eta_1^*+R_1}+e^{\eta_1+\eta_2^*+\delta_0}+e^{\eta_2+\eta_1^*+\delta_0^*}+e^{\eta_2+\eta_2^*+R_2}+e^{2\eta_1+2\eta_1^*+\epsilon_{11}}\nonumber\\
\hspace{-2.5cm}&&+e^{2\eta_1+2\eta_2^*+\epsilon_{12}}+e^{2\eta_2+2\eta_1^*+\epsilon_{21}}+e^{2\eta_2+2\eta_2^*+\epsilon_{22}}+e^{2\eta_1+\eta_1^*+\eta_2^*+\tau_1}+e^{2\eta_1^*+\eta_1+\eta_2+\tau_1^*}\nonumber\\
\hspace{-2.5cm}&& +e^{2\eta_2+\eta_1^*+\eta_2^*+\tau_2}+e^{2\eta_2^*+\eta_1+\eta_2+\tau_2^*}+e^{\eta_1+\eta_1^*+\eta_2+\eta_2^*+R_3}+e^{2\eta_1+2\eta_1^*+\eta_2+\eta_2^*+\theta_{11}}\nonumber\\
\hspace{-2.5cm}&&
+e^{2\eta_1+2\eta_2^*+\eta_2+\eta_1^*+\theta_{12}}+e^{2\eta_2+2\eta_1^*+\eta_1+\eta_2^*+\theta_{21}}+e^{2\eta_2+2\eta_2^*+\eta_1+\eta_1^*+\theta_{22}}+e^{2(\eta_1+\eta_1^*+\eta_2+\eta_2^*)+R_4},~~~~~~
\eea
and the auxiliary function $s$ is given by
\bea
\hspace{-2.5cm}s=&& ((\alpha_1^{(1)})^2+(\alpha_1^{(2)})^2) e^{2 \eta_1}+((\alpha_2^{(1)})^2+(\alpha_2^{(2)})^2) e^{2 \eta _2}+2(\alpha_1^{(1)} \alpha_2^{(1)}+\alpha_1^{(2)} \alpha_2^{(2)})e^{\eta _1+\eta _2}\nonumber\\
\hspace{-2.5cm}&&+e^{\eta _1+\eta _1^*+2\eta _2+\lambda _{11}}+e^{\eta _1+\eta _2^*+2 \eta _2+\lambda _{12}}+e^{\eta _2+\eta _1^*+2 \eta_1+\lambda _{21}}+e^{\eta _2+\eta _2^*+2 \eta _1+\lambda _{22}} \nonumber\\
\hspace{-2.5cm}&&+e^{2 \eta _1+2 \eta _1^*+2 \eta _2+\lambda _1}+e^{2 \eta _1+2 \eta _2+2 \eta _2^*+\lambda _2}+e^{2\eta _1+\eta_1^*+2 \eta _2+\eta _2^*+\lambda _3}.
\eea
\label{2s2c}\endnumparts
Here, $\eta_l=k_l(t+i k_l z)$, $l=1,2$. Various other quantities appearing in the above equation (10) can be obtained from the Appendix by substituting $m=2$. The above general two-soliton solution is characterized by six complex parameters $k_1,~k_2,~\alpha_1^{(1)},~\alpha_2^{(1)},~\alpha_1^{(2)}$ and $\alpha_2^{(2)}$.

\section{Bright soliton solutions of 3-component CCNLS equations}
We obtain the exact bright one- and two-soliton solutions of three component CCNLS equations in this section by applying the non-standard Hirota's bilinearization method described in section 2.
\subsection{\textbf{Bright (3,1) soliton solution}}
The bright (3,1) soliton solution of CCNLS equation (\ref{e2nc}) with $m=3$, can be obtained by terminating the power series expansion (6) as $g^{(j)}=\chi g_1^{(j)}+\chi^3 g_3^{(j)}$, $j=1,2,3$, $f=1+\chi^2 f_2+\chi^4 f_4$, $s=\chi^2 s_2$ and by solving the set of partial differential equations arising at like powers of $\chi$. Then the (3,1) soliton solution can be written as
\numparts
\bea
q_j&=&\frac{\alpha_1^{(j)} e^{\eta_1}+e^{2\eta_1+\eta_1^*+\delta_{11}^{(j)}}}{1+e^{\eta_1+\eta_1^*+R_1}+e^{2\eta_1+2\eta_1^*+\epsilon_{11}}},\quad ~j=1,2,3, \label{os3cc1}
\eea

\noindent where
\bea
e^{\delta_{11}^{(j)}}=\frac{\gamma \alpha_1^{(j)*} \displaystyle\sum_{l=1}^3 (\alpha_1^{(l)})^2}{2 (k_1+k_1^*)^2},~ \quad
e^{R_1}=\frac{\kappa_{11}}{(k_1+k_1^*)}, \quad j=1,2,3,\\
e^{\epsilon_{11}}=\frac{\gamma^2 \Big|\displaystyle\sum_{j=1}^3 {(\alpha_1^{(j)})^2}\Big|^2}{4 (k_1+k_1^*)^4}, \quad \kappa_{11}=\frac{\gamma \displaystyle\sum_{j=1}^3 {|\alpha_1^{(j)}|^2}}{(k_1+k_1^*)}.~~
\eea
The auxiliary function $s$ is found to be
\bea
s&=&\sum_{j=1}^3 (\alpha_1^{(j)})^2 e^{2\eta_1}.
\eea \label{1s3c}\endnumparts
The above bright one-soliton solution can also be classified into the following ICS and CCS as in the previous section.\\

\begin{figure}[h]
\centering\includegraphics[width=0.56\linewidth]{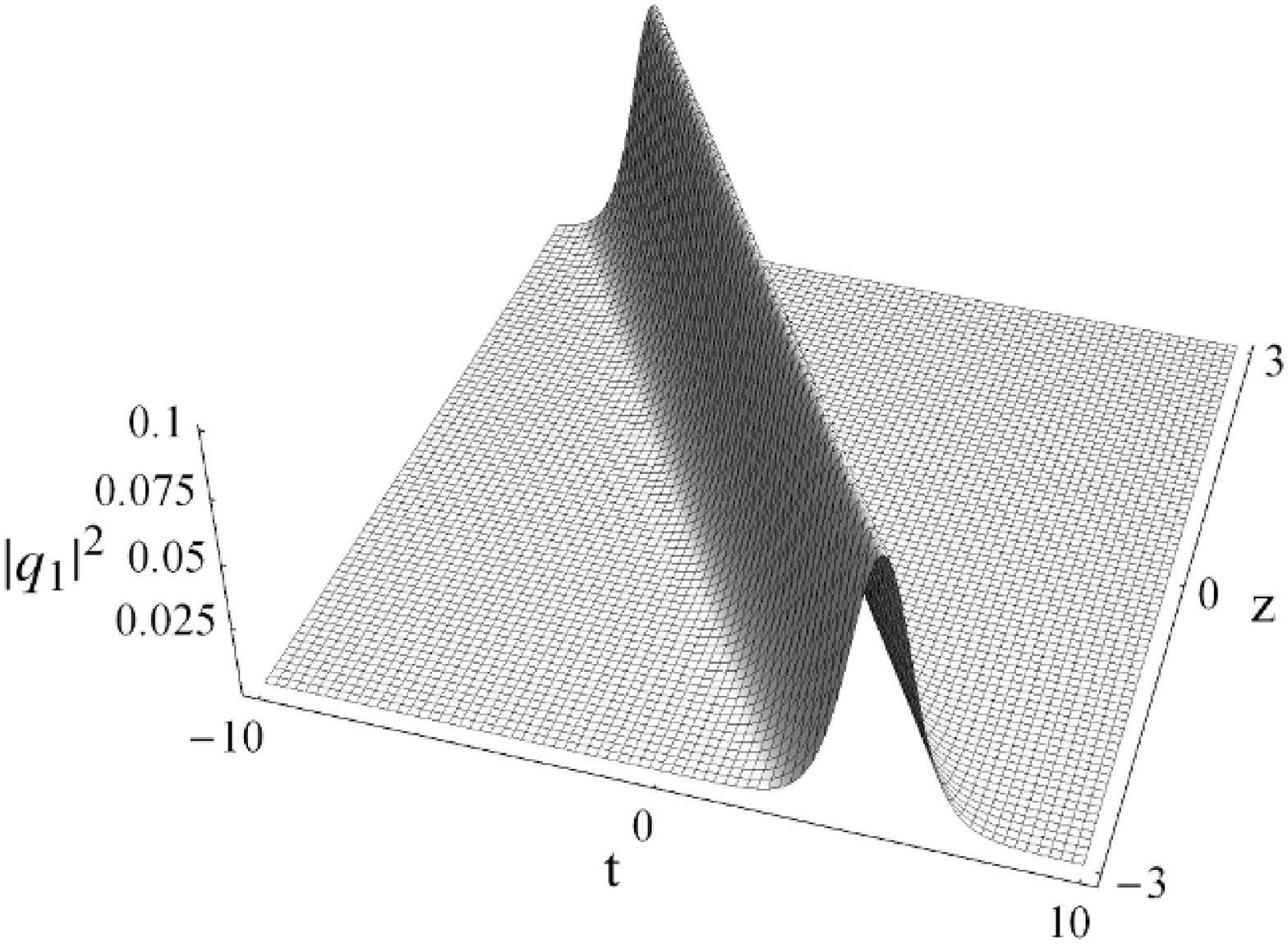}~\\~\includegraphics[width=0.56\linewidth]{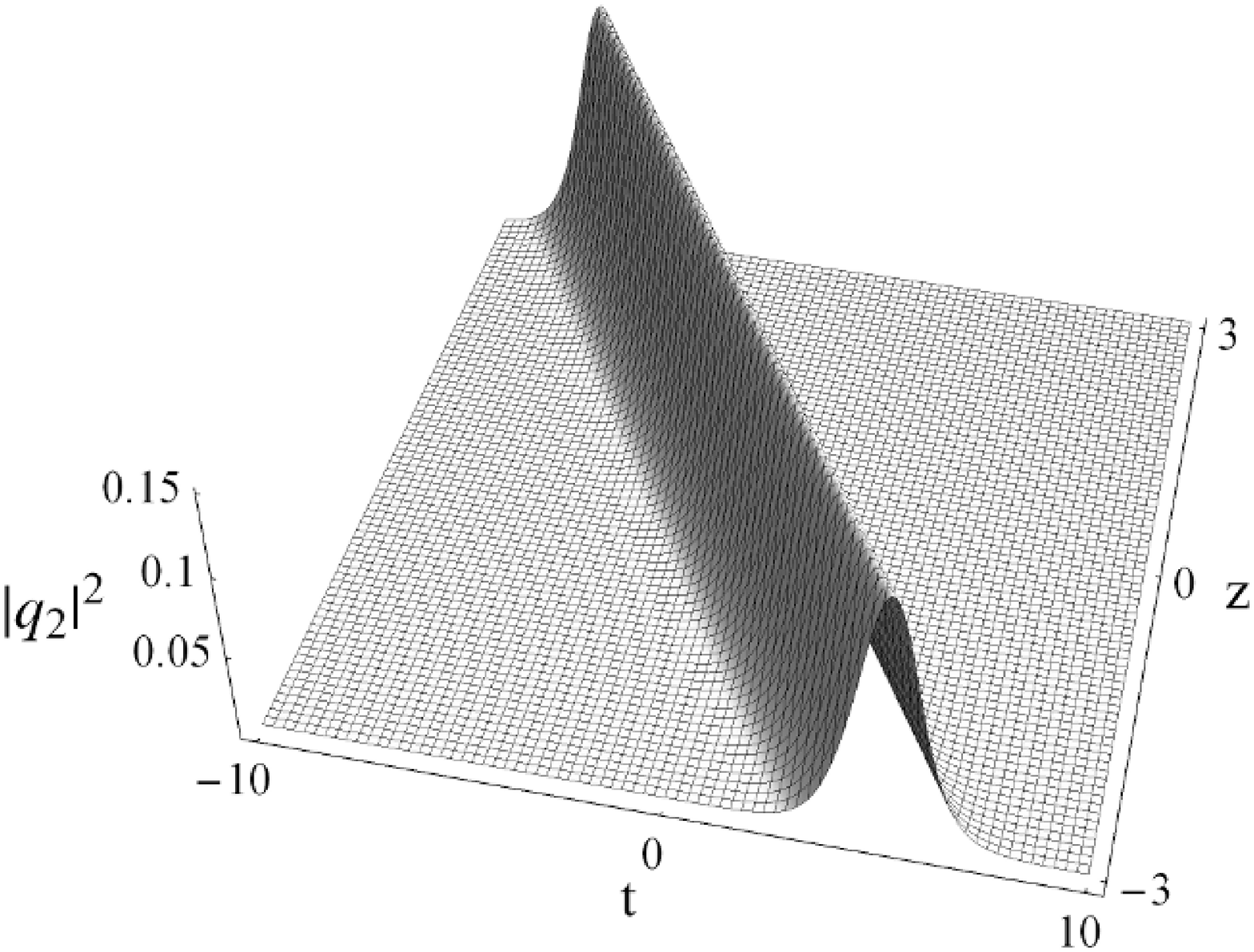}\\\includegraphics[width=0.56\linewidth]{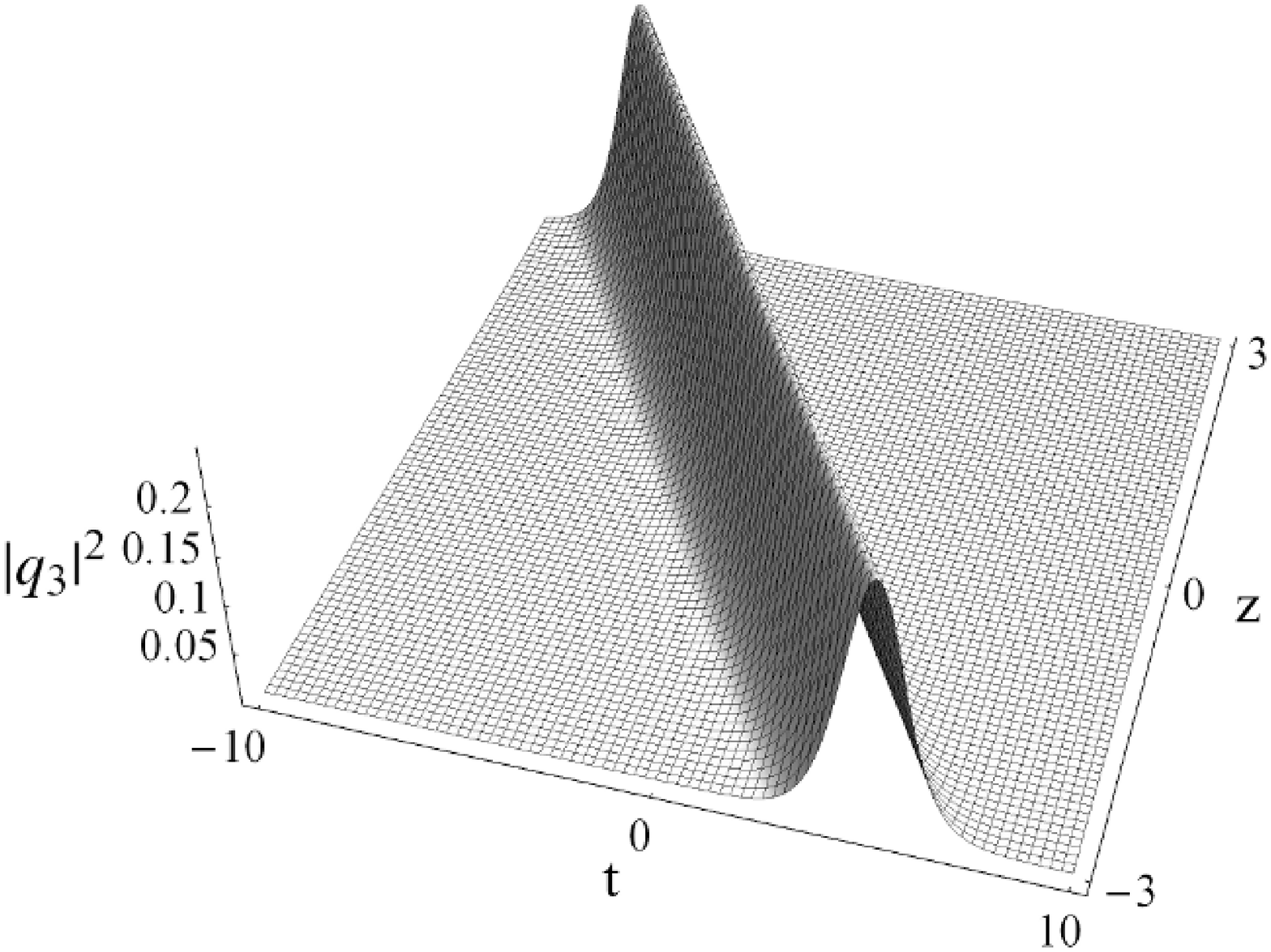}
\caption{Non-degenerate type (3,1) incoherently coupled soliton.}
\label{3cos1}
\end{figure}

\begin{figure}[h]
\centering\includegraphics[width=0.7\linewidth]{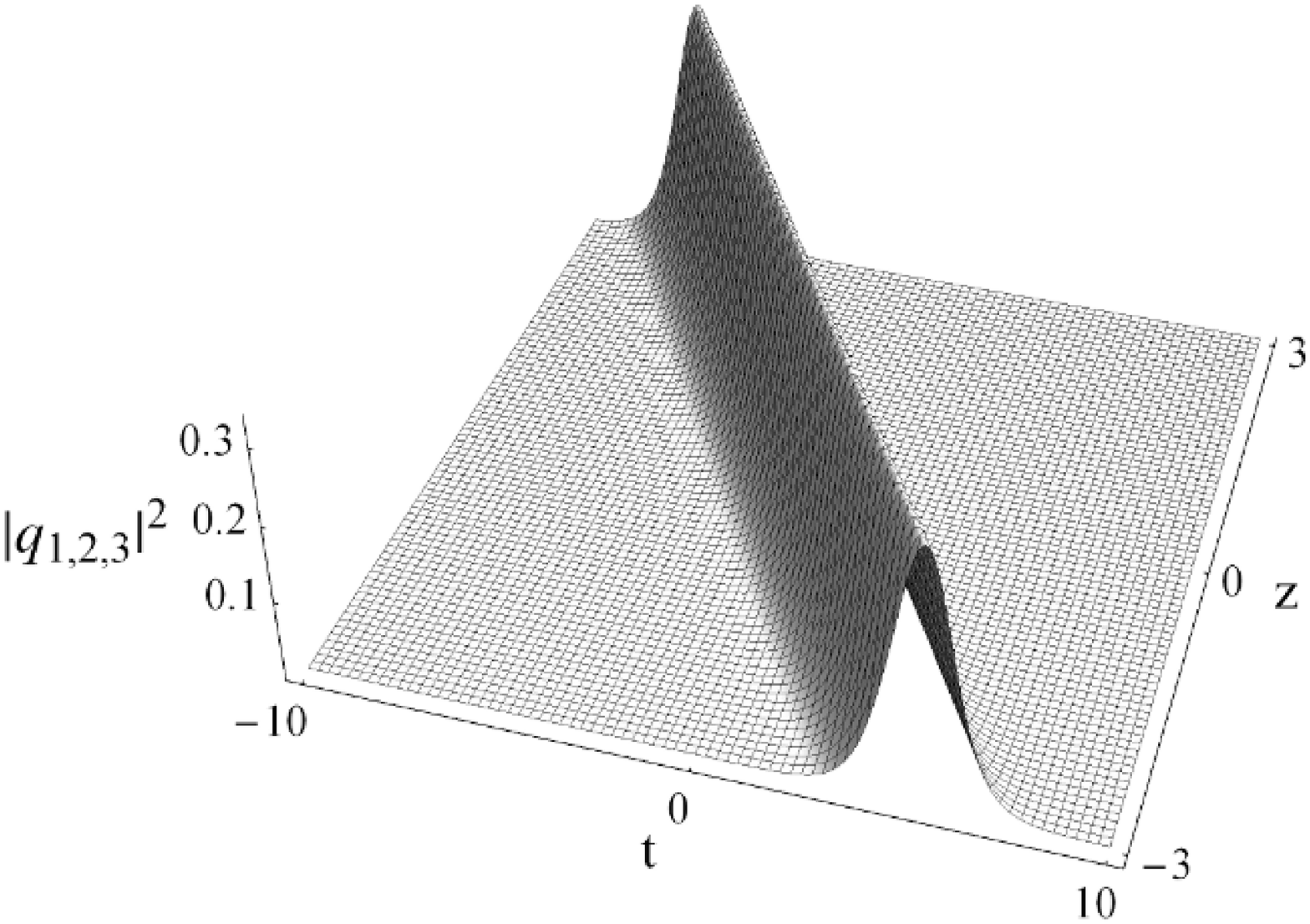}
\caption{Degenerate type (3,1) coherently coupled soliton.}
\label{3cos2}
\end{figure}

\begin{figure}[h]
\centering\includegraphics[width=0.56\linewidth]{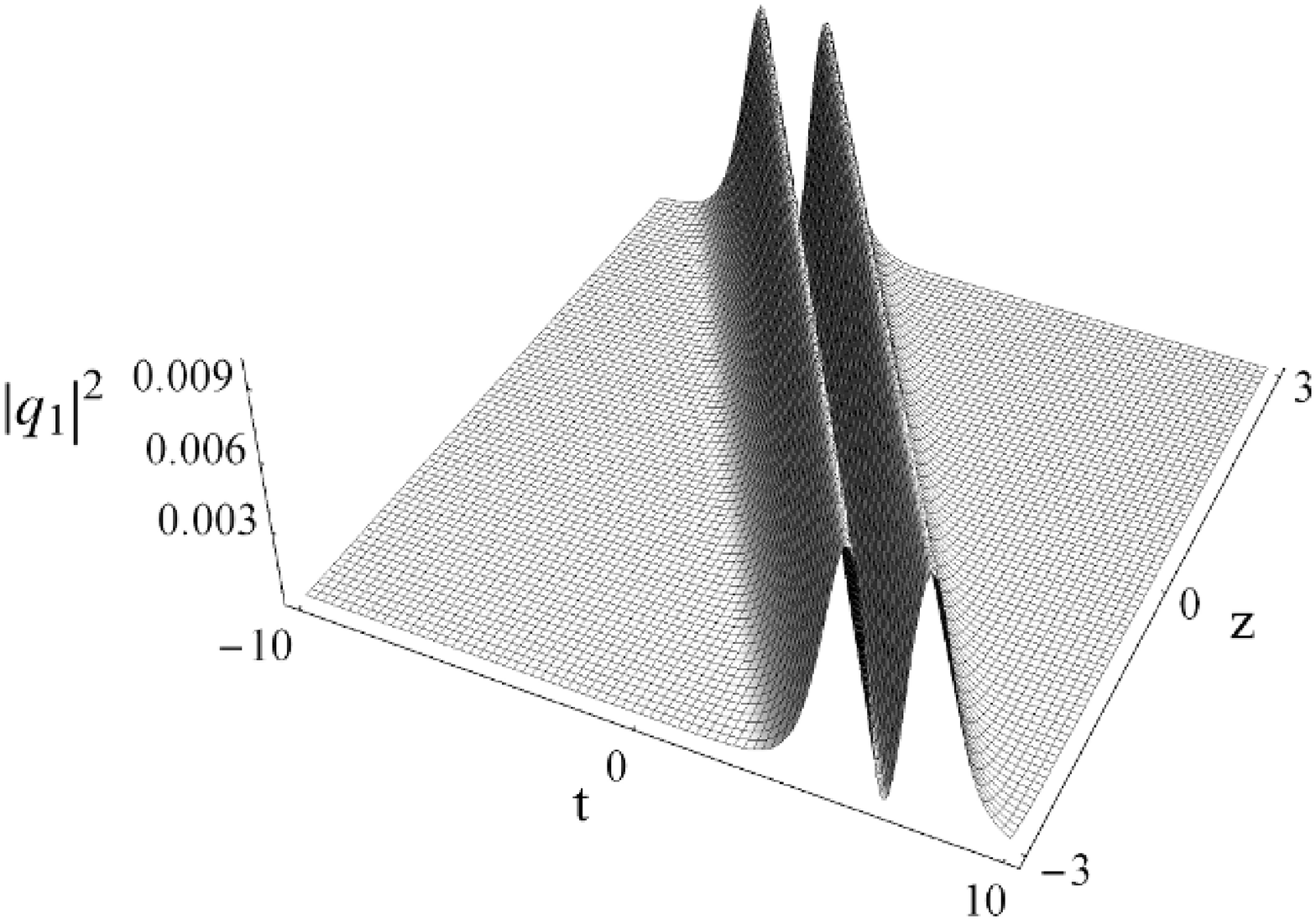}\\\centering\includegraphics[width=0.56\linewidth]{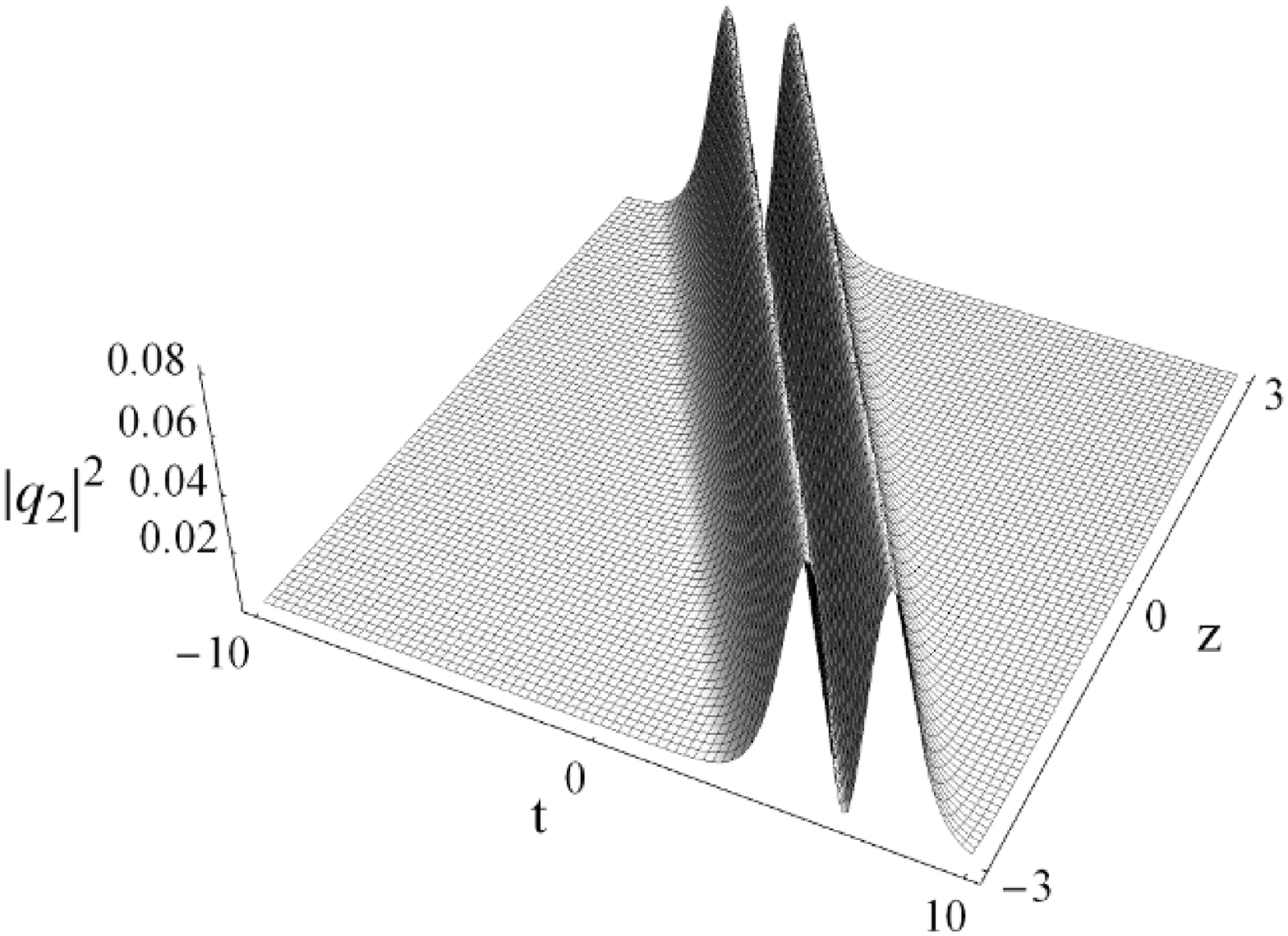}\\\includegraphics[width=0.56\linewidth]{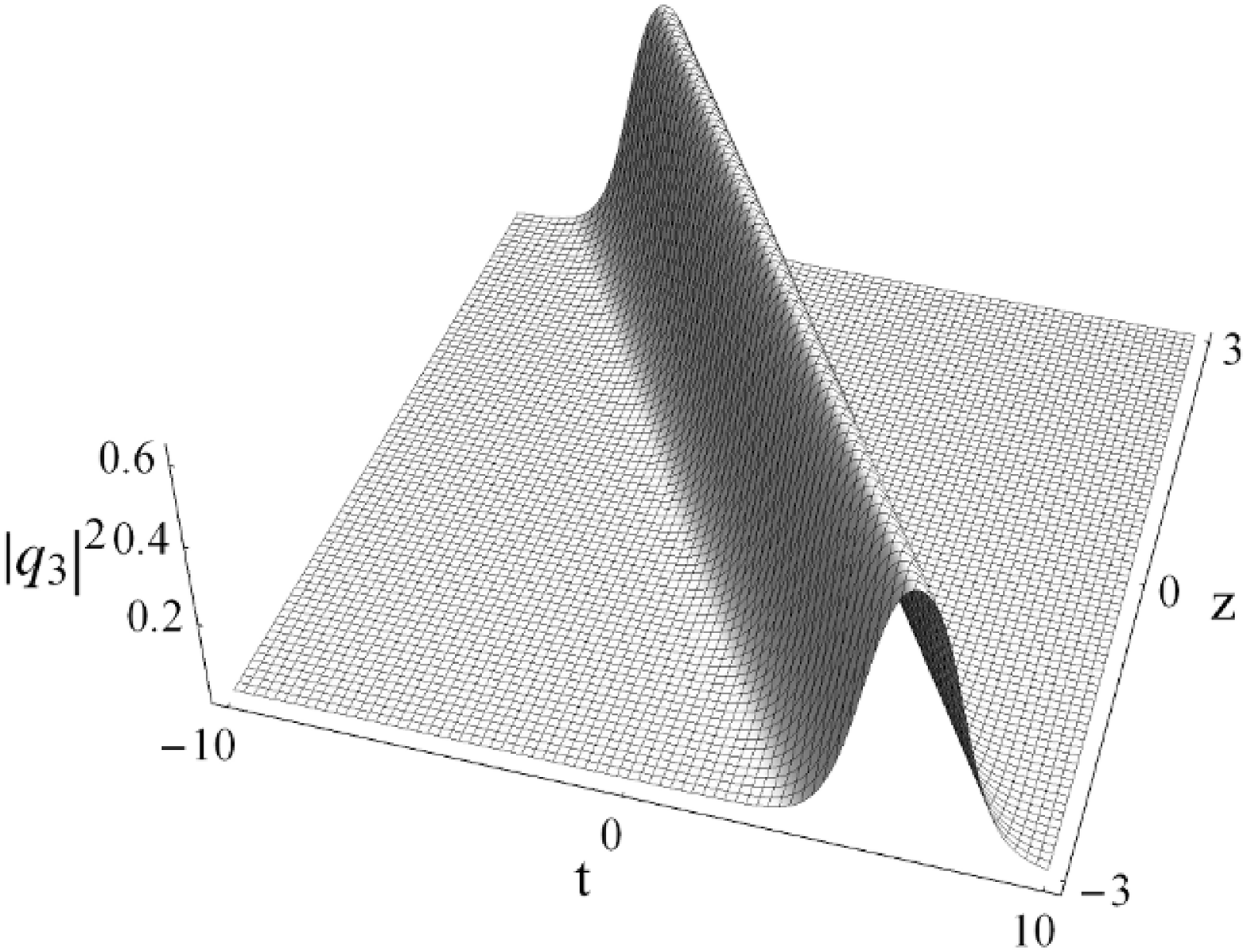}
\caption{Non-degenerate type (3,1) coherently coupled soliton.}
\label{3cos2a}
\end{figure}

\noindent (i) \underline{Bright (3,1) ICS:} \\
The (3,1) ICS appears for the choice $(\alpha_1^{(1)})^2+(\alpha_1^{(2)})^2+(\alpha_1^{(3)})^2 = 0$, and in its explicit form it reads as
\bea
q_j&=& {A_j}~\mbox{sech}\left(\eta_{1R}+\frac{R_1}{2} \right)e^{i\eta_{1I}}, \quad j=1,2,3,
\label{ms}
\eea
where $A_j=\alpha_1^{(j)} \sqrt{\frac{k_{1R}}{2 \kappa_{11}}}$, $R_1=\mbox{ln}\left(\frac{\kappa_{11}}{2k_{1R}}\right)$, $\eta_{1R}=k_{1R}(t-2k_{1I}z)$, and $\eta_{1I}=k_{1I}t+(k_{1R}^2-k_{1I}^2)z$. Unlike in the two-component case, here the degenerate ICS is not at all possible. These (3,1) ICSs admit either non-degenerate (completely different intensity profiles in all the three components) or partially degenerate (same intensity profiles in any two of the components) soliton profiles. These solitons are equivalent to the $3$-component Manakov type solitons \cite{man} and are characterized by three arbitrary complex parameters. Such non-degenerate type ICS arising for the parametric choice $k_1=1-i$, $\gamma=2$, $\alpha_1^{(1)}=\sqrt{2}$, $\alpha_1^{(2)}=\sqrt{3}$, and $\alpha_1^{(3)}=\sqrt{5}i$ is shown in figure \ref{3cos1}.
\\
\noindent (ii) \underline{Bright (3,1) CCS:} \\
The (3,1) CCSs appear for the choice $(\alpha_1^{(1)})^2+(\alpha_1^{(2)})^2+(\alpha_1^{(3)})^2\neq0$. The exact form of this soliton can be obtained by rewriting the one-soliton solution (\ref{os3cc1}) as
{\small\bea
q_j=2A_j\left(\frac{\mbox{cos}(P_j)~\mbox{cosh}\left(Q\right)
+i~\mbox{sin}(P_j)~\mbox{sinh}(Q)}{4 \mbox{cosh}^2(Q)+L}\right)e^{i\eta_{1I}}, \quad j=1,2,3,
\label{ccs3c}
\eea}
where $A_j=e^{{\frac{l_j+\delta_{11}^{(j)}-\epsilon_{11}}{2}}}$, $P_j=\frac{\delta_{11I}^{(j)}-l_{jI}}{2}$, $l_{j}=\ln({\alpha_1^{(j)}})$, $j=1, 2,3$, $Q=\eta_{1R}+\frac{\epsilon_{11}}{4}$, $L=e^{({R_1-\frac{\epsilon_{11}}{2}})}-2$, $\eta_{1R}=k_{1R}(t-2k_{1I}z)$, and $\eta_{1I}=k_{1I}t+(k_{1R}^2-k_{1I}^2)z$. These (3,1) CCSs admit both single hump and double hump profiles. In fact, we obtain perfect `sech' type (3,1) CCSs for specific choice of parameters satisfying the relation $\frac{\alpha_1^{(1)}}{\alpha_1^{(1)*}}=\frac{\alpha_1^{(2)}}{\alpha_1^{(2)*}}=\frac{\alpha_1^{(3)}}{\alpha_1^{(3)*}}$, which ultimately makes $P_j=L=0,~j=1,2,3,$ in the above equation (\ref{ccs3c}). The CCSs can also have degenerate intensity profiles in addition to non-degenerate profiles. A typical degenerate and also a non-degenerate type CCSs are shown in figure \ref{3cos2} and figure \ref{3cos2a} for the parametric choices $k_1=1-i$, $\gamma=2$, $\alpha_1^{(1)}=2$, $\alpha_1^{(2)}=2$, and $\alpha_1^{(3)}=2$ and $k_1=1-i$, $\gamma=2$, $\alpha_1^{(1)}=0.25$, $\alpha_1^{(2)}=-0.71$, and $\alpha_1^{(3)}=1.25i$, respectively.


\subsection{\textbf{Bright (3,2) soliton solution}}
The bright (3,2) soliton solution of system (\ref{e2nc}) with $m=3$, can be obtained as in the two-component case by restricting the power series expansion (6) and by allowing $j$ to run from 1 to 3. The two-soliton solution is found to be
\numparts\bea
q_j=\frac{g^{(j)}}{f},\quad j=1,2,3,
\eea
where
\bea
\hspace{-2.5cm}g^{(j)}=&&\alpha _1^{(j)} e^{\eta _1} +\alpha _2^{(j)} e^{\eta _2}+e^{2 \eta _1+\eta _1^{*}+\delta_{11}^{(j)}}+e^{2 \eta _1+\eta _2^{*}+\delta_{12}^{(j)}}
+e^{2 \eta _2+\eta _1^{*}+\delta_{21}^{(j)}}+e^{2 \eta _2+\eta _2^{*}+\delta_{22}^{(j)}}\nonumber\\
\hspace{-2.5cm}&&+e^{\eta_1+\eta_1^*+\eta_2+\delta_1^{(j)}}+e^{\eta_2+\eta_2^*+\eta_1+\delta_2^{(j)}}+e^{2\eta_1+2\eta_1^*+\eta_2+\mu_{11}^{(j)}}+e^{2\eta_1+2\eta_2^*+\eta_2+\mu_{12}^{(j)}}\nonumber\\
\hspace{-2.5cm}&&+e^{2\eta_2+2\eta_1^*+\eta_1+\mu_{21}^{(j)}}+e^{2\eta_2+2\eta_2^*+\eta_1+\mu_{22}^{(j)}}+e^{2\eta_1+\eta_1^*+\eta_2+\eta_2^*+\mu_1^{(j)}}\nonumber\\
\hspace{-2.5cm}&&+e^{2\eta_2+\eta_2^*+\eta_1+\eta_1^*+\mu_2^{(j)}}+e^{2\eta_1+2\eta_1^*+2\eta_2+\eta_2^*+\phi_1^{(j)}}+e^{2\eta_1+2\eta_2+2\eta_2^*+\eta_1^*+\phi_2^{(j)}},\quad j=1,2,3,\\
\hspace{-2.5cm}f=&&1+e^{\eta_1+\eta_1^*+R_1}+e^{\eta_1+\eta_2^*+\delta_0}+e^{\eta_2+\eta_1^*+\delta_0^*}+e^{\eta_2+\eta_2^*+R_2}+e^{2\eta_1+2\eta_1^*+\epsilon_{11}}\nonumber\\
\hspace{-2.5cm}&&+e^{2\eta_1+2\eta_2^*+\epsilon_{12}}+e^{2\eta_2+2\eta_1^*+\epsilon_{21}}+e^{2\eta_2+2\eta_2^*+\epsilon_{22}}+e^{2\eta_1+\eta_1^*+\eta_2^*+\tau_1}+e^{2\eta_1^*+\eta_1+\eta_2+\tau_1^*}\nonumber\\
\hspace{-2.5cm}&& +e^{2\eta_2+\eta_1^*+\eta_2^*+\tau_2}+e^{2\eta_2^*+\eta_1+\eta_2+\tau_2^*}+e^{\eta_1+\eta_1^*+\eta_2+\eta_2^*+R_3}+e^{2\eta_1+2\eta_1^*+\eta_2+\eta_2^*+\theta_{11}}\nonumber\\
\hspace{-2.5cm}&&
+e^{2\eta_1+2\eta_2^*+\eta_2+\eta_1^*+\theta_{12}}+e^{2\eta_2+2\eta_1^*+\eta_1+\eta_2^*+\theta_{21}}+e^{2\eta_2+2\eta_2^*+\eta_1+\eta_1^*+\theta_{22}}+e^{2(\eta_1+\eta_1^*+\eta_2+\eta_2^*)+R_4},~~~~~~~
\eea
and the auxiliary function
\bea
\hspace{-2.5cm}s=&& \sum_{j=1}^3 (\alpha_1^{(j)})^2 e^{2 \eta_1}+\sum_{j=1}^3 (\alpha_2^{(j)})^2 e^{2 \eta _2}+2\sum_{j=1}^3 (\alpha_1^{(j)} \alpha_2^{(j)})e^{\eta _1+\eta _2}+e^{\eta _1+\eta _1^*+2\eta _2+\lambda _{11}}\nonumber\\
\hspace{-2.5cm}&&+e^{\eta _1+\eta _2^*+2 \eta _2+\lambda _{12}}+e^{\eta _2+\eta _1^*+2 \eta_1+\lambda _{21}}+e^{\eta _2+\eta _2^*+2 \eta _1+\lambda _{22}}+e^{2 \eta _1+2 \eta _1^*+2 \eta _2+\lambda _1} \nonumber\\
\hspace{-2.5cm}&&+e^{2 \eta _1+2 \eta _2+2 \eta _2^*+\lambda _2}+e^{2\eta _1+\eta_1^*+2 \eta _2+\eta _2^*+\lambda _3}.
\eea
\label{2s3c}\endnumparts
In the above equations, $\eta_l=k_l(t+ik_l z),~l=1,2$, and the detailed expressions for other quantities can be obtained from the Appendix by substituting $m=3$. The above ($3,2$) solution is characterized by eight complex parameters.

\section{Soliton solutions of $m$--component CCNLS equations}
The bright one- and two-soliton solutions of arbitrary $m$-component system (\ref{e2nc}) can be obtained by generalizing the results of $2$-component and $3$-component cases. For completeness, we give the procedure to write down the multicomponent soliton solutions from $m=3$ case.
\subsection{\textbf{Bright ($m,1$) soliton solution}}
The ($m$,1) soliton solution can be obtained from (11) by allowing $j$ to run from 1 to $m$ and by redefining $\kappa_{11}$ and $e^{\epsilon_{11}}$ as $\kappa_{11}=\frac{\gamma}{(k_1+k_1^*)} \displaystyle\sum_{j=1}^m {|\alpha_1^{(j)}|^2}$ and $e^{\epsilon_{11}}=\frac{\gamma^2}{4 (k_1+k_1^*)^4} \big|\displaystyle\sum_{j=1}^m {(\alpha_1^{(j)})^2}\big|^2 $, respectively. It can also be found that the standard `sech' type ICS results for the choice $\displaystyle\sum_{j=1}^m (\alpha_1^{(j)})^2 = 0$ and the CCS arises for $\displaystyle\sum_{j=1}^m (\alpha_1^{(j)})^2 \neq 0$. One can also arrive at perfect `sech' type ($m$,1) CCS for the choice $\frac{\alpha_1^{(1)}}{\alpha_1^{(1)*}}=\frac{\alpha_1^{(2)}}{\alpha_1^{(2)*}}=...=\frac{\alpha_1^{(m)}}{\alpha_1^{(m)*}}$.

\subsection{\textbf{Bright ($m$,2) soliton solution}}
In a similar manner, the ($m$,2) soliton solution takes the form of (14) but with $j=1,2,3,...,m$  and all the other quantities appearing in the corresponding equations are defined in the Appendix.

\section{Soliton collisions in the multicomponent CCNLS equations}
The two-soliton solutions obtained in the preceding sections describe the interaction of two solitons in the multicomponent CCNLS system. One can get more insight into the collision dynamics and identify interesting collision properties by performing the asymptotic analysis of the two-soliton solution. In this section, we analyse the $2$-component and $3$-component cases and point out interesting behaviours which are not possible in the single component NLS equation. Our analysis can naturally be generalized to arbitrary $m$-component CCNLS system (\ref{e2nc}). To perform the asymptotic analysis of the two-soliton solution, we consider the parametric choices $k_{1R},~k_{2R}>0$ and $k_{1I}>k_{2I}$, without loss of generality. A similar analysis can also be carried out for other choices of parameters also.

\subsection{\textbf{Two-component CCNLS equations}}
The $2$-CCNLS equations can support both coherently coupled- and incoherently coupled-solitons as shown in the preceding sections. Now it is of interest to investigate the collision of a CCS with an ICS and also the collision between two CCSs and the collision among two ICSs, separately. These three types of soliton collisions in two-component CCNLS equations are discussed in this sub-section.

\subsubsection{Collision of coherently coupled soliton with incoherently coupled soliton:}
\indent First we consider the collision of a CCS, say $S_1$, with an ICS, say $S_2$, in which the former arises for the choice  $(\alpha_1^{(1)})^2+(\alpha_1^{(2)})^2 \neq 0$ and the latter results for the choice $(\alpha_2^{(1)})^2+(\alpha_2^{(2)})^2 = 0$. The asymptotic forms of the above solitons $S_1$ and $S_2$ before interaction ($z\rightarrow -\infty$) and after interaction ($z\rightarrow +\infty$) can be expressed as below. In the following equations, $\eta_l=k_l(t+ik_lz)$, $\eta_{lR}=k_{lR}(t-2k_{lI}z)$, and $\eta_{lI}=k_{lI}t+(k_{lR}^2-k_{lI}^2)z$, $l=1,2$.\\
\noindent{{\bf Before collision} ($z\rightarrow -\infty$)}\\
\underline{CCS $S_1$} ($\eta_{1R}\simeq0, \eta_{2R}\rightarrow -\infty$):
\numparts\bea
\hspace{-1.5cm}q_j^{1-}=2{A_j^{1-}} \left(\frac{\mbox{cos}(P_j^{1-})\mbox{cosh}(\eta_{1R}^-)+i~\mbox{sin}(P_j^{1-})\mbox{sinh}(\eta_{1R}^-)}{4\mbox{cosh}^2(\eta_{1R}^-)+L^{1-}}\right) e^{i\eta_{1I}}, \quad j=1,2,
\eea
where $A_j^{1-}= e^{\frac{\delta_{11}^{(j)}+l_j^--\epsilon_{11}}{2}}$, $P_j^{1-}= {\frac{\delta_{11I}^{(j)}-l_{jI}^-}{2}}$, $l_j^-=\mbox{ln}(\alpha_1^{(j)})$, $L^{1-}= e^{(R_1-\frac{\epsilon_{11}}{2})}-2$, and $\eta_{1R}^-=\eta_{1R}+\frac{\epsilon_{11}}{4}$.\\
\underline{ICS $S_2$} ($\eta_{2R}\simeq0, \eta_{1R}\rightarrow \infty$):
\bea
\hspace{-1.5cm}q_j^{2-}=\frac{A_j^{2-}}{2} \mbox{sech}\left(\eta_{2R}+\frac{\theta_{11}-\epsilon_{11}}{2}\right)e^{i\eta_{2I}}, \quad j=1,2,
\eea
where $A_j^{2-}=e^{\mu_{11}^{(j)}-\left(\frac{\epsilon_{11}+\theta_{11}}{2}\right)}$. Here and in the following, the superscript and subscript appearing in the quantities $q,~A,~P,~l,$ and $L$ represent the soliton number and the component, respectively. The sign $- ~(+)$ appearing in the superscript represents the asymptotic form of the soliton before (after) interaction.\\
\noindent{{\bf After collision} ($z\rightarrow +\infty$)}\\
\underline{CCS $S_1$} ($\eta_{1R}\simeq0, \eta_{2R}\rightarrow \infty$):
\bea
\hspace{-1.5cm}q_j^{1+}=2{A_j^{1+}} \left(\frac{\mbox{cos}(P_j^{1+})\mbox{cosh}(\eta_{1R}^+)+i~\mbox{sin}(P_j^{1+})\mbox{sinh}(\eta_{1R}^+)}{4\mbox{cosh}^2(\eta_{1R}^+)+L^{1+}}\right) e^{i\eta_{1I}}, \quad j=1,2,
\eea
where $A_j^{1+}= e^{(\frac{\mu_1^{(j)}+\delta_2^{(j)}-\theta_{11}-R_2}{2})}$, $P_j^{1+}= {\frac{\mu_{1I}^{(j)}-\delta_{2I}^{(j)}}{2}}$, $L^{1+}= e^{R_3-(\frac{\theta_{11}+R_2}{2})}-2$, and $\eta_{1R}^+=\eta_{1R}+\frac{\theta_{11}-R_2}{4}$.\\
\underline{ICS $S_2$} ($\eta_{2R}\simeq0, \eta_{1R}\rightarrow -\infty$):
\bea
\hspace{-1.5cm}q_j^{2+}=\frac{A_j^{2+}}{2} \mbox{sech}\left(\eta_{2R}+\frac{R_2}{2}\right)e^{i\eta_{2I}}, \quad j=1,2,
\eea\label{2col1}
\endnumparts
where $A_j^{2+}=\alpha_2^{(j)}e^{-(\frac{R_2}{2})}$. Explicit expressions for various quantities appearing in equation (15) can be deduced from the Appendix for $m=2$.

From the above asymptotic expressions (15), we can relate the amplitude (peak value of the envelope) of the CCS $S_1$ after interaction to that of before interaction as $A_j^{1+}=T_j~A_j^{1-},~j=1,2$, where the transition amplitudes $T_j$'s are given by
\bea
\hspace{-1.5cm}T_j=\left(\frac{(k_1^*+k_2)(k_1-k_2)\big|(\alpha_1^{(j)} \kappa_{22}-\alpha_2^{(j)} \kappa_{12})+\alpha_2^{(j)*} \Omega\big|^2}{(k_1+k_2^*) (k_1^*-k_2^*) ~\kappa_{22}^2 ~ |\alpha_1^{(j)}|^2}\right)^{\frac{1}{2}}, ~~ j=1,2,
\label{trans2ca}\eea
where $\Omega=\frac{\gamma(\alpha_1^{(1)}\alpha_2^{(1)}+\alpha_1^{(2)}\alpha_2^{(2)}) }{(k_1-k_2)}$. A careful analysis of equation (\ref{trans2ca}) shows that the absolute values of $T_1$ and $T_2$ will become one only if $(\alpha_1^{(1)})^2+(\alpha_1^{(2)})^2=0$, for which one can not have CCS. Hence we conclude that in the 2-component CCNLS system (\ref{e2}), during collision between CCS and an ICS, CCS always experiences an intensity/energy switching. In addition to the above amplitude variation, CCS $S_1$ experiences an amplitude dependent phase shift $\Phi_1=\frac{\theta_{11}-R_2-\epsilon_{11}}{4k_{1R}} \equiv \frac{1}{k_{1R}}\ln\left(\frac{(k_1-k_2)(k_1^*-k_2^*)}{(k_1+k_2^*)(k_1^*+k_2)}\right)$. On the other hand, the ICS undergoes elastic collision as the amplitude of the ICS $S_2$ before and after collision can be related through the expression $A_j^{2+}=\frac{(k_1^* + k_2)(k_1^* - k_2^*)}{(k_1 - k_2)(k_1 + k_2^*)}~A_j^{2-},$ and hence $|A_j^{2+}|^2=|A_j^{2-}|^2,~j=1,2$. Also, ICS $S_2$ undergoes a phase shift $\Phi_2=\frac{R_2+\epsilon_{11}-\theta_{11}}{2k_{2R}} \equiv -\left(\frac{2k_{1R}}{k_{2R}}\right)\Phi_1$. The change in the relative separation distance between these two solitons can be written as
\bea
\Delta t_{12}=t_{12}^- - t_{12}^+ = \left(1+\frac{2 k_{1R}}{k_{2R}}\right) \Phi_1,
\eea
where $t_{12}^-=t_{2}^- - t_{1}^-$ and $t_{12}^+=t_{2}^+ - t_{1}^+$, in which $t_{l}^-$ ($t_{l}^+$) is the position of soliton $S_l$ before (after) collision. Here the phase shifts and the relative separation distance between the solitons purely depend on $k$'s but they are  independent of $\alpha$-parameters.

\begin{figure}[h]
\centering
\includegraphics[width=0.65\linewidth]{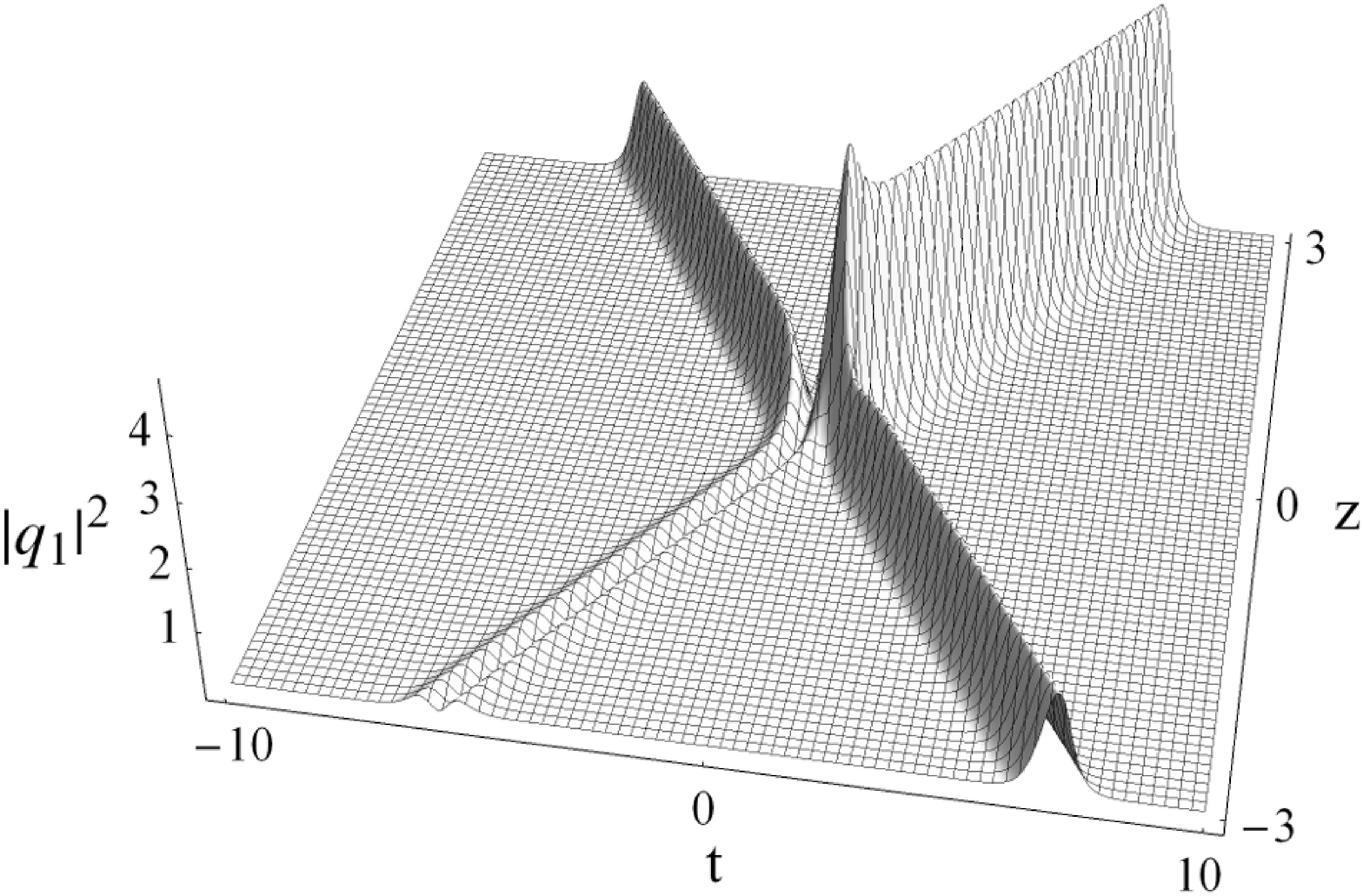}~\\~\includegraphics[width=0.65\linewidth]{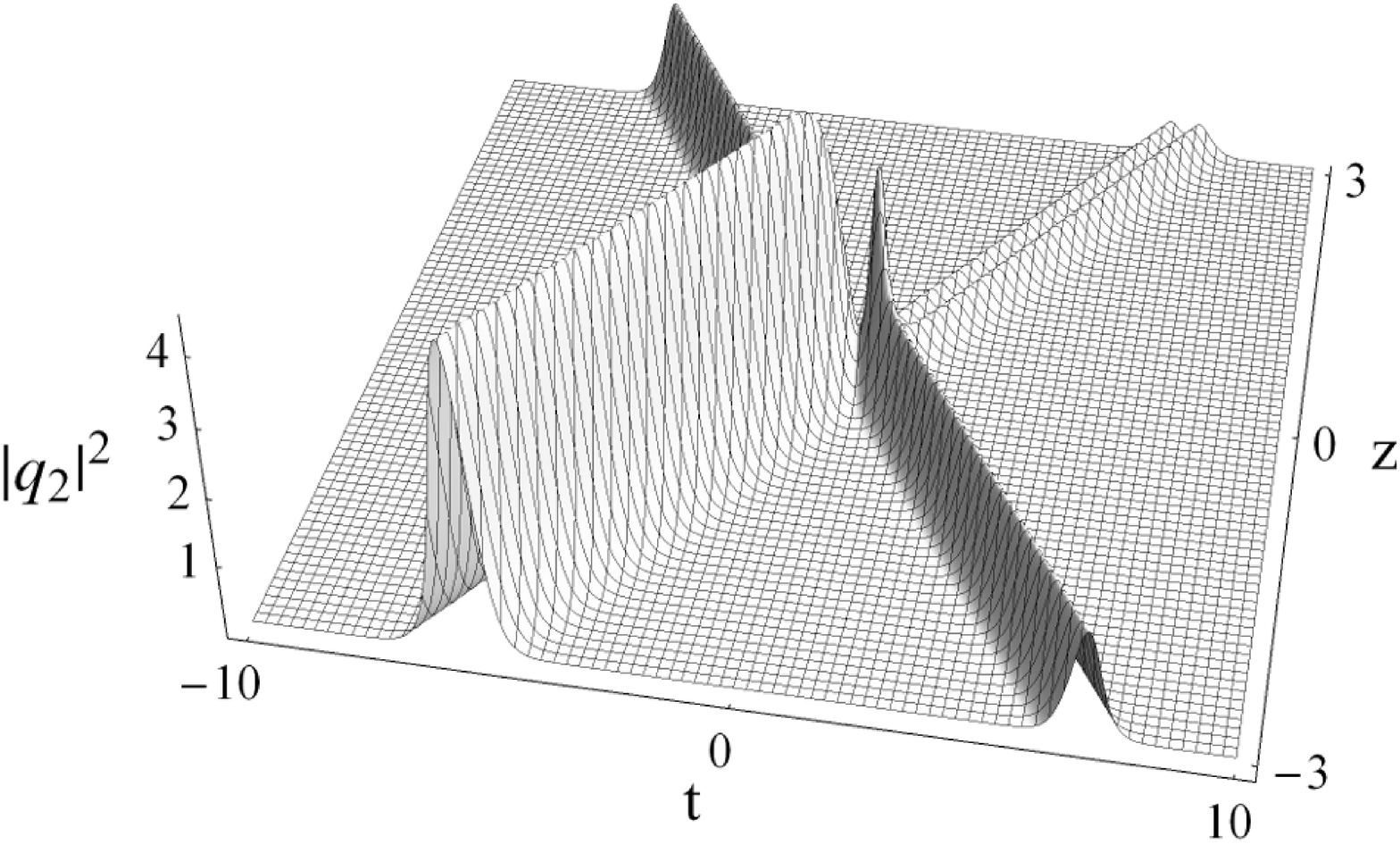}\\
\caption{Energy switching collision of a coherently coupled soliton $S_1$ with an incoherently coupled soliton $S_2$ in $2$-CCNLS system.}
\label{2cc1}
\end{figure}

The main physics behind this collision scenario is that during collision always there is an energy/intensity switching in the CCS but the ICS remains unaffected, thereby leading to energy non-conservation in individual components. However, the total energy is conserved. This follows directly from the following expressions
\numparts\bea
i~\frac{d}{dz}\int_{-\infty}^{\infty}|q_1|^2~dt = \gamma \int_{-\infty}^{\infty} (q_1^{*2}q_2^2-q_1^2q_2^{*2})~dt,\\
i~\frac{d}{dz}\int_{-\infty}^{\infty}|q_2|^2~dt = \gamma \int_{-\infty}^{\infty} (q_1^2q_2^{*2}-q_1^{*2}q_2^2)~dt,
\eea\endnumparts
which can be obtained from (\ref{e2}) in a straightforward way. In fact, the above relation shows that the energy of the individual component, ie. $\int_{-\infty}^{\infty}|q_j|^2~dt$, $j=1,2$, is not conserved whereas the total energy, that is, $\int_{-\infty}^{\infty}(|q_1|^2+|q_2|^2)~dt$, is conserved. As a consequence of this, the ICS induces significant energy switching in the CCS with an amplitude dependent phase shift and reappears elastically after interaction. During collision CCS can also switch its profile from single hump to double hump and vice-versa. This type of energy switching collision resulting in a dramatic alteration of soliton intensity profile is quite different from the shape changing collision of the Manakov solitons \cite{{tkprl},{rk}} and is shown in figure \ref{2cc1}, for illustrative purpose. The parameters are chosen as $k_1=2.3+i, ~k_2=2.5-i, ~\gamma=2,~ \alpha_1^{(1)}=0.75i,~ \alpha_1^{(2)}=1.9,~ \alpha_2^{(1)}=1+i$ and $\alpha_2^{(2)}=1-i$. Here, the non-degenerate CCS $S_1$ switches its profile from double (single) hump to a single (double) hump and exhibits enhancement (suppression) of intensity in the $q_1$ ($q_2$) component, whereas the degenerate ICS $S_2$ undergoes mere elastic collision.

\subsubsection{Collision of coherently coupled solitons:}
Let us consider the collision of two coherently coupled solitons, arising for the choice $(\alpha_j^{(1)})^2+(\alpha_j^{(2)})^2 \neq 0,~j=1,2$. Their asymptotic forms are found as\\
\noindent{{\bf Before collision} ($z\rightarrow -\infty$)}\\
\underline{CCS $S_1$}:
\numparts\bea
\hspace{-1.5cm}q_j^{1-}={2A_j^{1-}} \left(\frac{\mbox{cos}(P_j^{1-})\mbox{cosh}(\eta_{1R}^-)+i~\mbox{sin}(P_j^{1-})\mbox{sinh}(\eta_{1R}^-)}{ 4\mbox{cosh}^2(\eta_{1R}^-)+L^{1-}}\right) e^{i\eta_{1I}}, \quad j=1,2,
\label{cca}\eea
where $A_j^{1-}= e^{\frac{\delta_{11}^{(j)}+l_j^--\epsilon_{11}}{2}}$, $P_j^{1-}= {\frac{\delta_{11I}^{(j)}-l_{jI}^-}{2}}$, $l_j^-=\mbox{ln}(\alpha_1^{(j)})$, $j=1,2$, $L^{1-}=e^{(R_1-\frac{\epsilon_{11}}{2})}-2$, and $\eta_{1R}^-=\eta_{1R}+\frac{\epsilon_{11}}{4}$.\\
\noindent\underline{CCS $S_2$}:
\bea
\hspace{-1.5cm}q_j^{2-}={2A_j^{2-}} \left(\frac{\mbox{cos}(P_j^{2-})\mbox{cosh}(\eta_{2R}^-)+i~\mbox{sin}(P_j^{2-})\mbox{sinh}(\eta_{2R}^-)}{ 4\mbox{cosh}^2(\eta_{2R}^-)+L^{2-}}\right) e^{i\eta_{2I}}, \quad j=1,2,
\eea
where $A_j^{2-}= e^{\frac{\phi_1^{(j)}+\mu_{11}^{(j)}-R_4-\epsilon_{11}}{2}}$, $P_j^{2-}= {\frac{\phi_{1I}^{(j)}-\mu_{11I}^{(j)}}{2}}$, $L^{2-}=e^{\theta_{11}-(\frac{R_4+\epsilon_{11}}{2})}-2$, and $\eta_{2R}^-=\eta_{2R}+\frac{R_4-\epsilon_{11}}{4}$.\\
\noindent{{\bf After collision} ($z\rightarrow +\infty$)}\\
\underline{CCS $S_1$}:
\bea
\hspace{-1.5cm}q_j^{1+}={2A_j^{1+}} \left(\frac{\mbox{cos}(P_j^{1+})\mbox{cosh}(\eta_{1R}^+)+i~\mbox{sin}(P_j^{1+})\mbox{sinh}(\eta_{1R}^+)}{4\mbox{cosh}^2(\eta_{1R}^+)+L^{1+}}\right) e^{i\eta_{1I}}, \quad j=1,2,
\eea
where $A_j^{1+}= \frac{(k_1-k_2)(k_1^*+k_2)}{(k_1^*-k_2^*)(k_1+k_2^*)}A_j^{1-}$, $P_j^{1+}= {\frac{\phi_{2I}^{(j)}-\mu_{22I}^{(j)}}{2}} \equiv P_j^{1-}$, $L^{1+}=e^{\theta_{22}-(\frac{R_4+\epsilon_{22}}{2})}-2 \equiv L^{1-}$, and $\eta_{1R}^+=\eta_{1R}+\frac{R_4-\epsilon_{22}}{4}$.\\
\underline{CCS $S_2$}:
\bea
\hspace{-1.5cm}q_j^{2+}=2A_j^{2+} \left(\frac{\mbox{cos}(P_j^{2+})\mbox{cosh}(\eta_{2R}^+)+i~\mbox{sin}(P_j^{2+})\mbox{sinh}(\eta_{2R}^+)}{ 4\mbox{cosh}^2(\eta_{2R}^+)+L^{2+}}\right) e^{i\eta_{2I}}, \quad j=1,2,
\label{ccd}\eea\label{2c2ccs}\endnumparts
where $A_j^{2+}= \frac{(k_1^*-k_2^*)(k_1^*+k_2)}{(k_1-k_2)(k_1+k_2^*)}A_j^{2-}$, $P_j^{2+}= {\frac{\delta_{22I}^{(j)}-\l_{jI}^+}{2}} \equiv P_j^{2-}$, $l_j^+=\mbox{ln}(\alpha_2^{(j)})$, $L^{2+}=e^{(R_2-\frac{\epsilon_{22}}{2})}-2 \equiv L^{2-}$, and $\eta_{2R}^+=\eta_{2R}+\frac{\epsilon_{22}}{4}$. The other quantities appearing in the above equation (19) can be obtained from the Appendix for $m=2$.

\begin{figure}[t]
\centering
\includegraphics[width=0.65\linewidth]{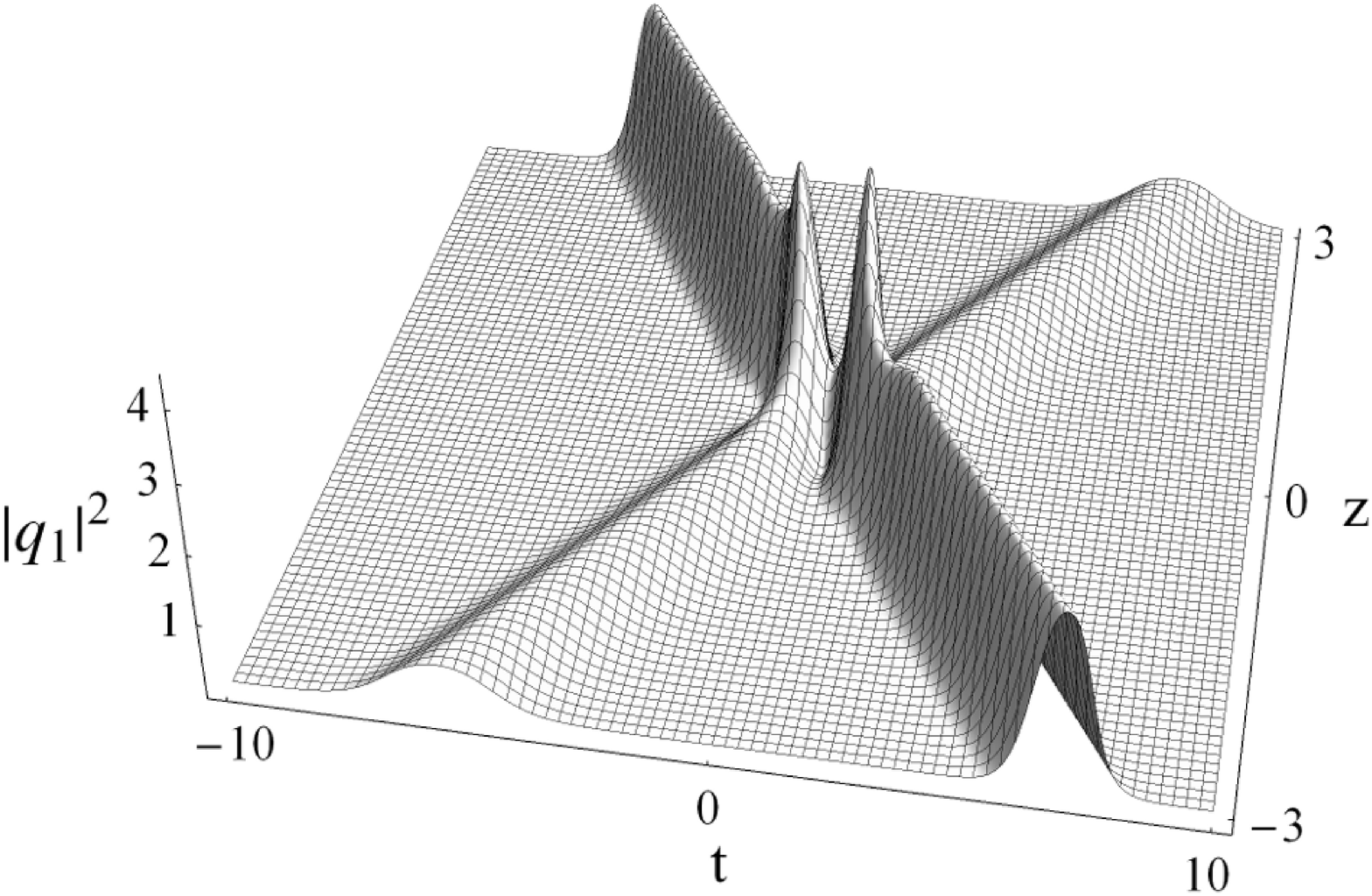}~\\~\includegraphics[width=0.65\linewidth]{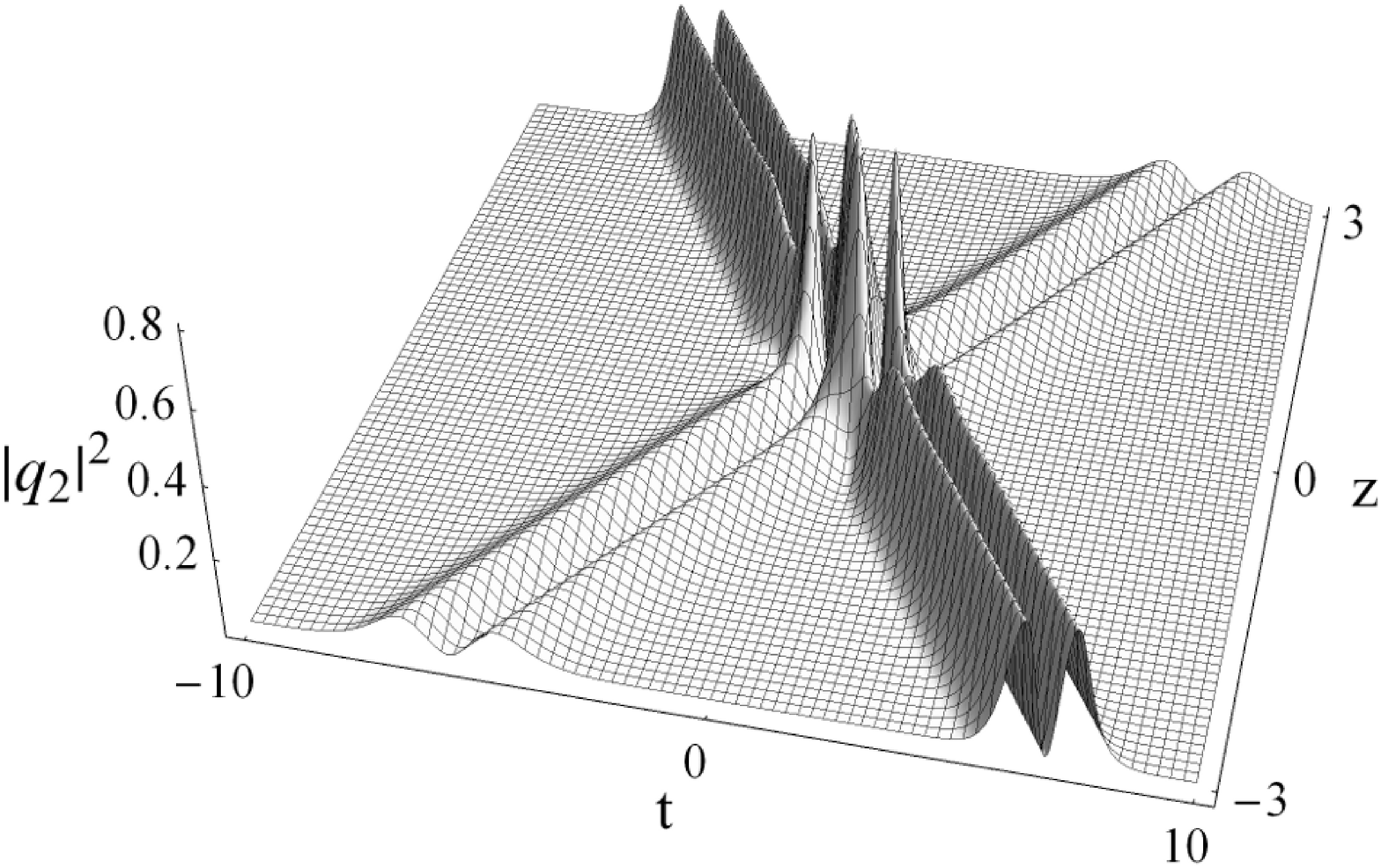}
\caption{Elastic collision of coherently coupled solitons in $2$-CCNLS system.}
\label{2cc2}
\end{figure}

From the above equations (\ref{cca})--(\ref{ccd}) we find $|A_j^{i+}|^2=|A_j^{i-}|^2,~i,j=1,2$. This clearly indicates that soliton intensities remain same before and after collision. Hence the CCSs undergo elastic collision with amplitude dependent phase shift, $\Phi_1=\frac{R_4-\epsilon_{11}-\epsilon_{22}}{4k_{1R}} \equiv \frac{1}{k_{1R}}\ln\left(\frac{(k_1-k_2)(k_1^*-k_2^*)}{(k_1+k_2^*)(k_1^*+k_2)}\right)$ for soliton $S_1$, and $\Phi_2=-\left(\frac{k_{1R}}{k_{2R}}\right)\Phi_1$ for soliton $S_2$. Here the change in the relative separation distance between the two CCSs is $\Delta t_{12}= \left(1+\frac{k_{1R}}{k_{2R}}\right) \Phi_1$. Such type of elastic collision of the CCSs is depicted in figure \ref{2cc2} for $k_1=1.5+i, ~k_2=2-i, ~\gamma=2,~ \alpha_1^{(1)}=1.7i,~ \alpha_1^{(2)}=1,~ \alpha_2^{(1)}=2i$ and $\alpha_2^{(2)}=1.2$. In figure \ref{2cc2}, two CCSs having single hump profiles in $q_1$ component and double hump profiles in $q_2$ component undergo elastic collision in both the components.

\subsubsection{Collision of incoherently coupled solitons:}
In this subsection, we discuss the collision dynamics of two ICSs arising for the choices $(\alpha_j^{(1)})^2+(\alpha_j^{(2)})^2 = 0,~j=1,2$. The following expressions are the asymptotic forms of solitons $S_1$ and $S_2$ before and after interaction.\\
 \noindent{{\bf Before collision} ($z\rightarrow -\infty$)}
\numparts\bea
\hspace{-2.7cm}\left(\begin{array}{c}
  q_j^{1-} \\
  q_j^{2-}
\end{array}\right)^{\bf T} = \frac{1}{2}
    \left(\begin{array}{c}
         A_j^{1-} e^{i\eta_{1I}}\\
        A_j^{2-}e^{i\eta_{2I}}
     \end{array} \right)^{\bf T}
     \left(\begin{array}{cc}
         \mbox{sech}\left(\eta_{1R}+\frac{R_1}{2}\right)~~ & 0 \\
         ~0 & \mbox{sech}\left(\eta_{2R}+\frac{R_3-R_1}{2}\right) \\
       \end{array}\right),
j=1,2,~~~~~~
\eea
where $A_j^{1-}=\alpha_1^{(j)} e^{-\frac{R_1}{2}}$ and $A_j^{2-}=e^{\delta_1^{(j)}-(\frac{R_1+R_3}{2})}$.\\
\noindent{{\bf After collision} ($z\rightarrow +\infty$)}
\bea
\hspace{-2.7cm}\left(\begin{array}{c}
  q_j^{1+} \\
  q_j^{2+}
\end{array}\right)^{\bf T}  = \frac{1}{2}
    \left(\begin{array}{c}
         A_j^{1+} e^{i\eta_{1I}}\\
        A_j^{2+}e^{i\eta_{2I}}
     \end{array} \right)^{\bf T}
     \left(\begin{array}{cc}
         \mbox{sech}\left(\eta_{1R}+\frac{R_3-R_2}{2}\right)~~ & 0 \\
         ~0 & \mbox{sech}\left(\eta_{2R}+\frac{R_2}{2}\right) \\
       \end{array}\right),
j=1,2,~~~~~~
\eea\label{2c2ics}\endnumparts
where $A_j^{1+}=e^{\delta_2^{(j)}-(\frac{R_2+R_3}{2})}$, $A_j^{2+}=\alpha_2^{(j)} e^{-\frac{R_2}{2}}$,  ``${\bf T} $" denotes the transpose of the matrix and various other quantities can be obtained from the Appendix for $m=2$. In the above equation (20) the superscripts  $1$ and $2$ appearing in $q_j$ and $A_j$ denote the solitons while $j$ denotes the component.

\begin{figure}[h]
\centering
\includegraphics[width=0.65\linewidth]{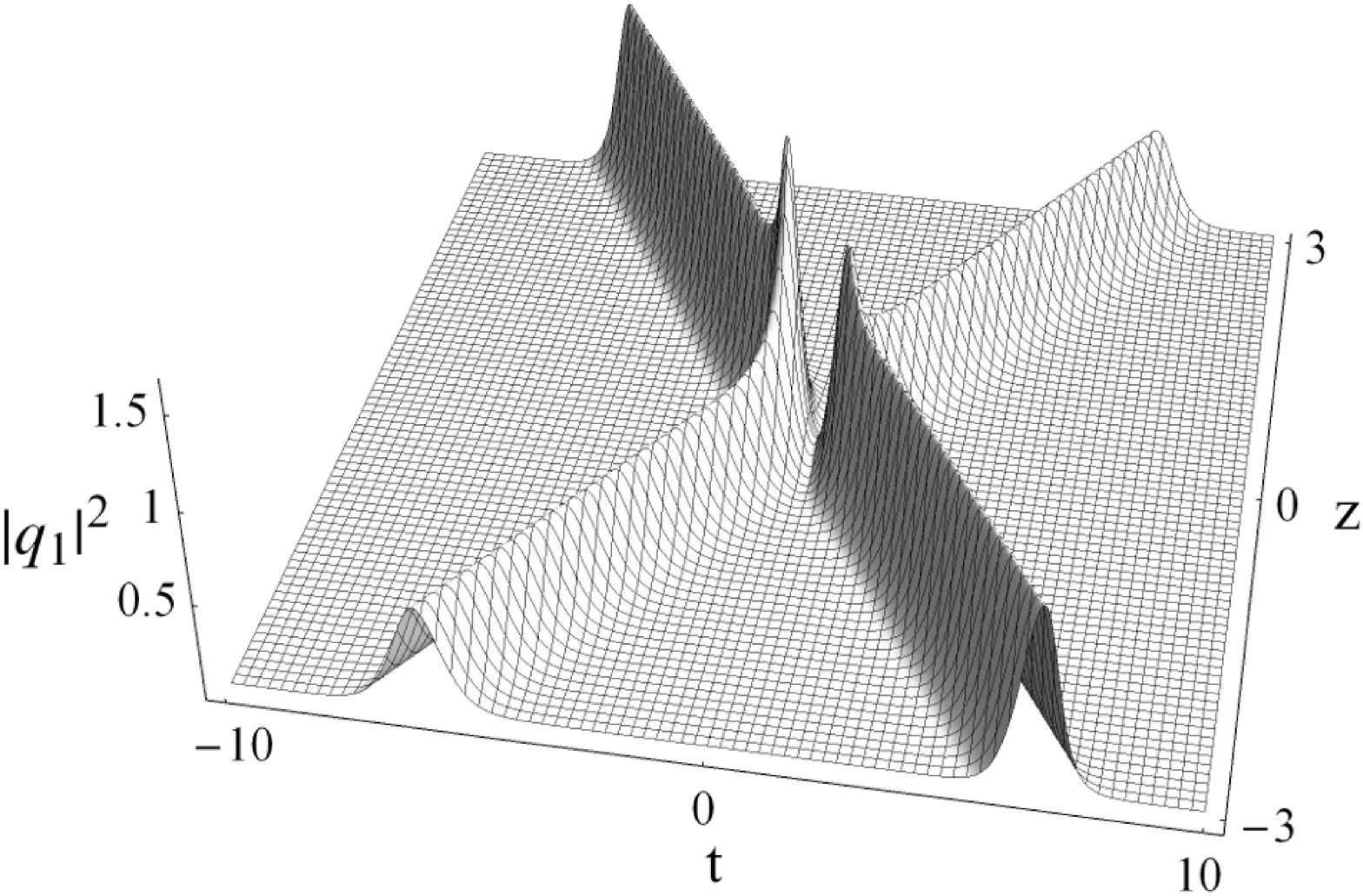}~\\~\includegraphics[width=0.65\linewidth]{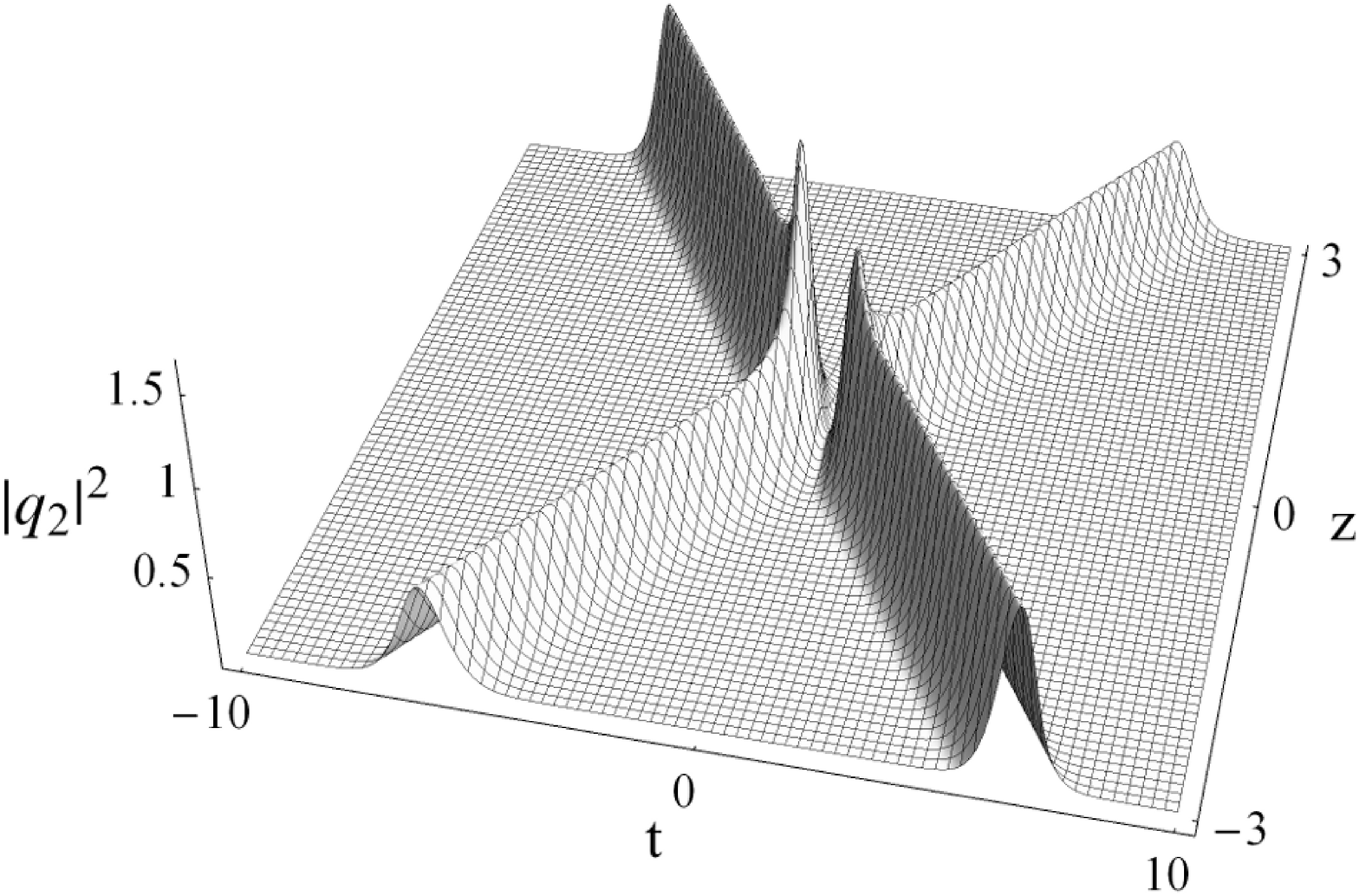}
\caption{Elastic collision of incoherently coupled solitons in $2$-CCNLS system.}
\label{2c2icsf}
\end{figure}

The amplitudes of the ICSs $S_1$ and $S_2$ after collision can be written from the above asymptotic expressions as $A_j^{1+}=\frac{(k_1-k_2)(k_1^*+k_2)}{(k_1^*-k_2^*)(k_1+k_2^*)}A_j^{1-}$ and $A_j^{2+}=\frac{(k_1^*-k_2^*)(k_1^*+k_2)}{(k_1-k_2)(k_1+k_2^*)} A_j^{2-}$, respectively. One can find the intensities of the solitons before and after collision are same, that is $|A_j^{l+}|^2=|A_j^{l-}|^2$, $j,l=1,2$. Thus the collision between the ICSs arising in the two-component case is always elastic. But $S_1$ and $S_2$ suffer phase-shifts $\Phi_1=\frac{R_3-R_2-R_1}{2k_{1R}}\equiv \frac{1}{k_{1R}}\ln\left(\frac{(k_1-k_2)(k_1^*-k_2^*)}{(k_1+k_2^*)(k_1^*+k_2)}\right)$ and $\Phi_2=\frac{R_1+R_2-R_3}{2k_{2R}}\equiv-\left(\frac{k_{1R}}{k_{2R}}\right)\Phi_1$, respectively, as observed from equation (20). From the above phase-shifts one can find the change in the relative separation distance as $\Delta t_{12}= \left(1+\frac{k_{1R}}{k_{2R}}\right) \Phi_1$. This type of elastic collision is exactly similar to the elastic collision of standard NLS solitons and is given in figure \ref{2c2icsf} for $k_1=1.5+i, ~k_2=2-i, ~\gamma=2,~ \alpha_1^{(1)}=1+i,~ \alpha_1^{(2)}=1-i,~ \alpha_2^{(1)}=1+2i$ and $\alpha_2^{(2)}=2-i$.

\subsection{\textbf{Three-component CCNLS equations}}
Next, we investigate the above three types of soliton collisions in the three-component CCNLS equations to examine how they differ from the $2$-component case.
\subsubsection{Collision of coherently coupled soliton with incoherently coupled soliton:}
Here we consider the collision between the coherently coupled soliton say $S_1$ ($(\alpha_1^{(1)})^2+(\alpha_1^{(2)})^2+(\alpha_1^{(3)})^2 \neq 0$) and an incoherently coupled soliton  $S_2$ ($(\alpha_2^{(1)})^2+(\alpha_2^{(2)})^2+(\alpha_2^{(3)})^2 = 0$). The asymptotic expressions of the CCS $S_1$ and ICS $S_2$ before and after interaction can be obtained as before and are given below explicitly.\\
\noindent{{\bf Before collision} ($z\rightarrow -\infty$)}\\
\underline{CCS $S_1$} ($\eta_{1R}\simeq0, \eta_{2R}\rightarrow -\infty$):
\numparts\bea
\hspace{-1.5cm}q_j^{1-}=2{A_j^{1-}} \left(\frac{\mbox{cos}(P_j^{1-})\mbox{cosh}(\eta_{1R}^-)+i~\mbox{sin}(P_j^{1-})\mbox{sinh}(\eta_{1R}^-)}{4\mbox{cosh}^2(\eta_{1R}^-)+L^{1-}}\right) e^{i\eta_{1I}}, \quad j=1,2,3,
\eea
where $A_j^{1-}= e^{\frac{\delta_{11}^{(j)}+l_j^--\epsilon_{11}}{2}}$, $P_j^{1-}= {\frac{\delta_{11I}^{(j)}-l_{jI}^-}{2}}$, $l_j^-=\mbox{ln}(\alpha_1^{(j)})$, $L^{1-}=e^{(R_1-\frac{\epsilon_{11}}{2})}-2$, and $\eta_{1R}^-=\eta_{1R}+\frac{\epsilon_{11}}{4}$.\\
\underline{ICS $S_2$} ($\eta_{2R}\simeq0, \eta_{1R}\rightarrow \infty$):
\bea
\hspace{-1.5cm}q_j^{2-}=\frac{A_j^{2-}}{2} \mbox{sech}\left(\eta_{2R}+\frac{\theta_{11}-\epsilon_{11}}{2}\right)e^{i\eta_{2I}}, \quad j=1,2,3,
\eea
where $A_j^{2-}=e^{\mu_{11}^{(j)}-\left(\frac{\epsilon_{11}+\theta_{11}}{2}\right)}$.\\
\noindent{{\bf After collision} ($z\rightarrow +\infty$)}\\
\underline{CCS $S_1$} ($\eta_{1R}\simeq0, \eta_{2R}\rightarrow \infty$):
\bea
\hspace{-1.5cm}q_j^{1+}=2{A_j^{1+}} \left(\frac{\mbox{cos}(P_j^{1+})\mbox{cosh}(\eta_{1R}^+)+i~\mbox{sin}(P_j^{1+})\mbox{sinh}(\eta_{1R}^+)}{4\mbox{cosh}^2(\eta_{1R}^+)+L^{1+}}\right) e^{i\eta_{1I}}, \quad j=1,2,3,
\eea
where $A_j^{1+}= e^{(\frac{\mu_1^{(j)}+\delta_2^{(j)}-\theta_{11}-R_2}{2})}$, $P_j^{1+}= {\frac{\mu_{1I}^{(j)}-\delta_{2I}^{(j)}}{2}}$, $L^{1+}=e^{R_3-(\frac{\theta_{11}+R_2}{2})}-2$, and $\eta_{1R}^+=\eta_{1R}+\frac{\theta_{11}-R_2}{4}$.\\
\underline{ICS $S_2$} ($\eta_{2R}\simeq0, \eta_{1R}\rightarrow -\infty$):
\bea
\hspace{-1.5cm}q_j^{2+}=\frac{A_j^{2+}}{2} \mbox{sech}\left(\eta_{2R}+\frac{R_2}{2}\right)e^{i\eta_{2I}}, \quad j=1,2,3,
\eea\label{ccics3c}
where $A_j^{2+}=\alpha_2^{(j)}e^{-\frac{R_2}{2}}$.

The amplitudes of the CCS $S_1$ before and after interaction can be related as $A_j^{1+}=T_j~A_j^{1-},~j=1,2,3$. Here the transition amplitudes $T_j$'s are defined as
\bea
\hspace{-1.5cm}T_j=\left(\frac{(k_1^*+k_2)(k_1-k_2)\big|(\alpha_1^{(j)} \kappa_{22}-\alpha_2^{(j)} \kappa_{12})+\alpha_2^{(j)*} \Omega\big|^2}{(k_1+k_2^*) (k_1^*-k_2^*) ~\kappa_{22}^2 ~ |\alpha_1^{(j)}|^2}\right)^{\frac{1}{2}}, ~~ j=1,2,3,
\label{tr3c}\eea  \endnumparts
where $\Omega=\frac{\gamma }{(k_1-k_2)}\displaystyle\sum_{l=1}^3 (\alpha_1^{(l)}\alpha_2^{(l)})$.

\begin{figure}[h]
\centering
\includegraphics[width=0.56\linewidth]{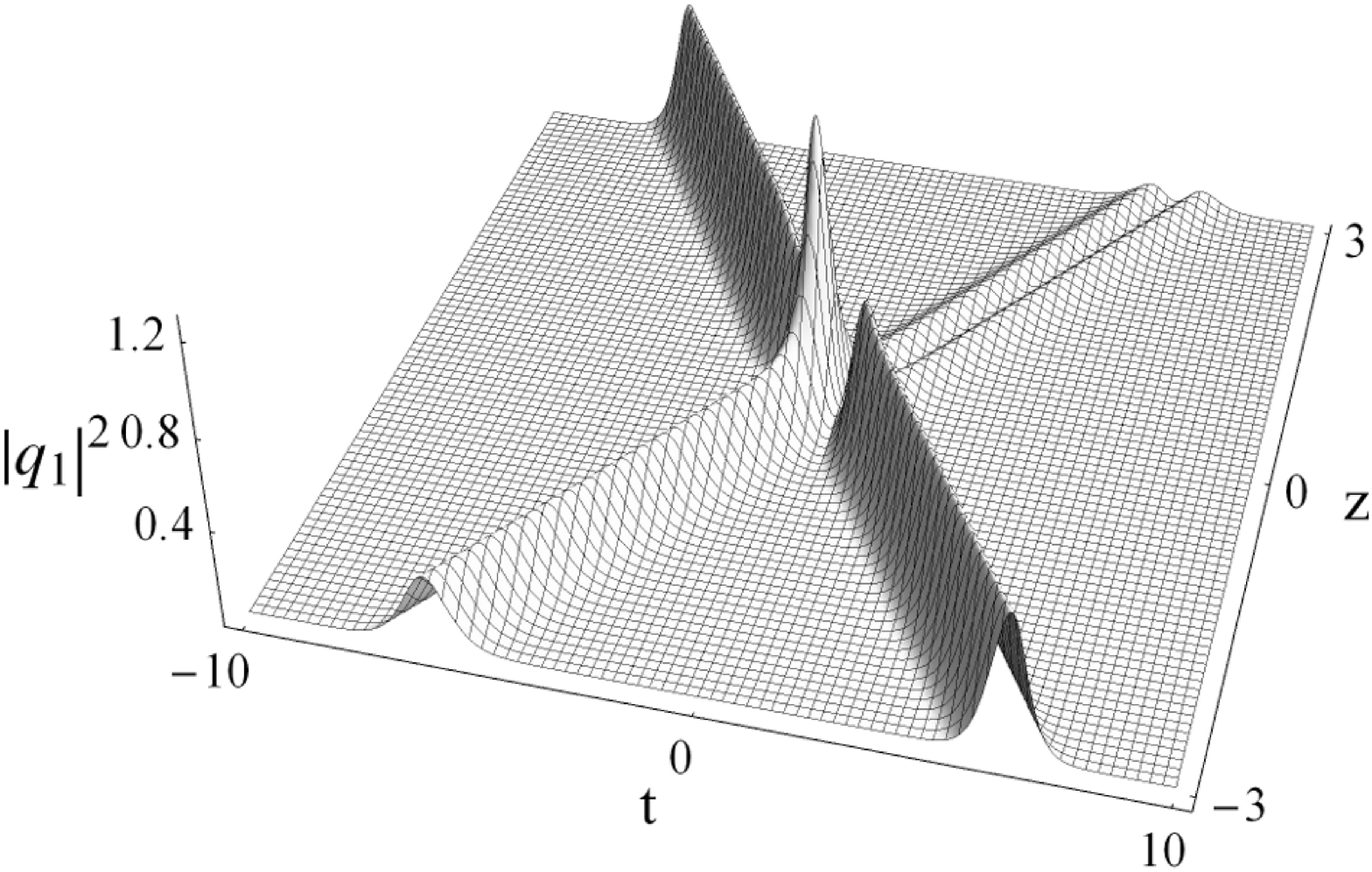}~\\~\includegraphics[width=0.56\linewidth]{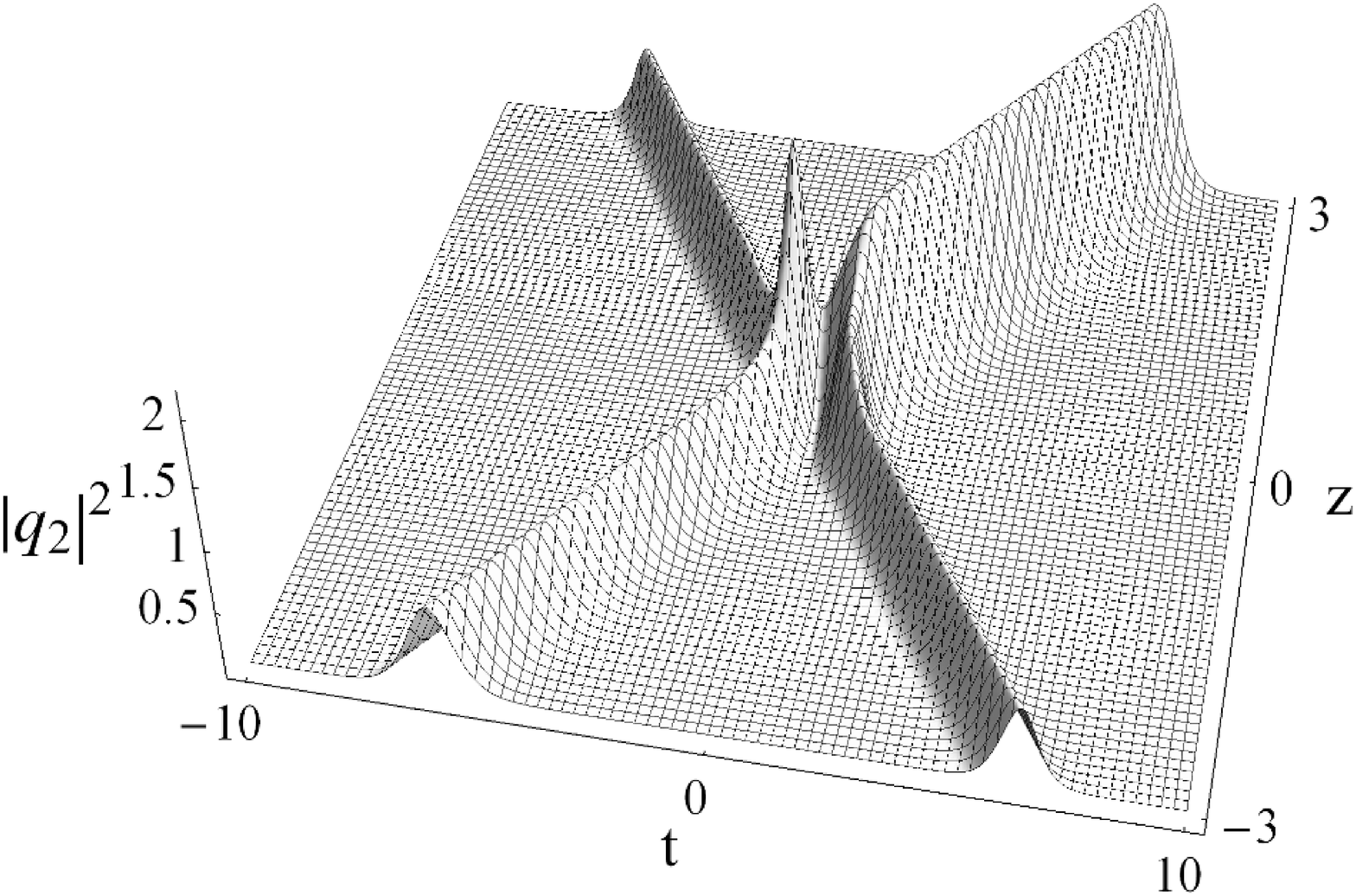}\\~~\includegraphics[width=0.56\linewidth]{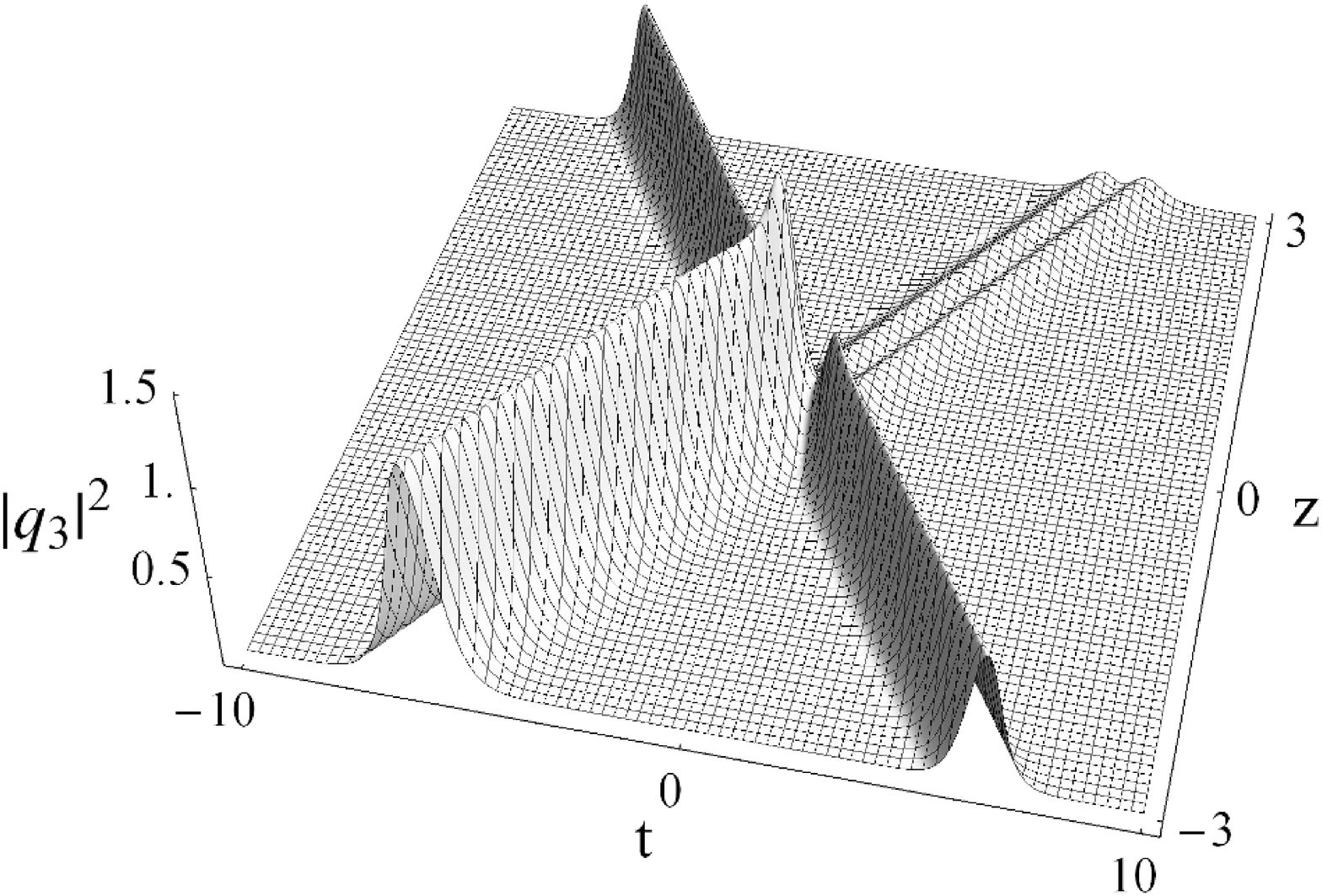}
\caption{Energy switching collision of a coherently coupled soliton with an incoherently coupled soliton in $3$-CCNLS system.}
\label{3cc1}
\end{figure}

One can check from equation (\ref{tr3c}) that the transition amplitudes become unimodular only for the choice $(\alpha_1^{(1)})^2+(\alpha_1^{(2)})^2+(\alpha_1^{(3)})^2 =0$, for which CCS can not exist. Thus the intensities of the colliding solitons before and after collision are always different and the CCS $S_1$ undergoes energy switching collision with shape alteration due to intensity switching among the components and amplitude dependent phase shift $\Phi_1=\frac{\theta_{11}-R_2-\epsilon_{11}}{4k_{1R}} \equiv \frac{1}{ k_{1R}}\ln \left(\frac{(k_1 - k_2)(k_1^* - k_2^*)}{(k_1^* + k_2)(k_1 + k_2^*)}\right)$. But the ICS $S_2$ reappears elastically after collision with the CCS and its amplitudes in the three components after collision can be expressed as $A_j^{2+}=\frac{(k_1^* + k_2)(k_1^* - k_2^*)}{(k_1 - k_2)(k_1 + k_2^*)}~A_j^{2-},~j=1,2,3,$ from which we find $|A_j^{2+}|^2=|A_j^{2-}|^2$. However $S_2$ suffers a phase shift $\Phi_2=\frac{R_2+\epsilon_{11}-\theta_{11}}{2k_{2R}} \equiv -\left(\frac{2k_{1R}}{k_{2R}}\right) \Phi_1$. The phase-shifts which are independent of $\alpha$-parameters also lead to a change in the relative separation distance of solitons $S_1$ and $S_2$ which is determined as $\Delta t_{12} = \left(1+\frac{2 k_{1R}}{k_{2R}}\right) \Phi_1$. This type of shape changing collision scenario is quite different from the shape changing (energy sharing) collision of three-component Manakov solitons \cite{{tkprl},{rk}} and is shown in figure \ref{3cc1} for the parametric choice $k_1=1.5+i, ~k_2=2-i, ~\gamma=2,~ \alpha_1^{(1)}=1,~ \alpha_1^{(2)}=1.5,~\alpha_1^{(3)}=2,~ \alpha_2^{(1)}=2+i$, $\alpha_2^{(2)}=2-i$, and $\alpha_2^{(3)}=\sqrt{6}~i$. Here, the CCS $S_1$ experiences suppression in its intensity after collision and also changes its profile from single hump to double hump in $q_1$ and $q_3$ components but in $q_2$ its intensity gets enhanced and also its single hump profile is retained. But the  intensity of ICS $S_2$ remains intact during collision in all the three components.
\subsubsection{Collision of coherently coupled solitons:}
In this subsection, we analyse the collision behaviour of two coherently coupled solitons arising in the three component CCNLS equations for the choice $(\alpha_j^{(1)})^2+(\alpha_j^{(2)})^2+(\alpha_j^{(3)})^2 \neq 0,~j=1,2$. The asymptotic expressions are given below.\\
\noindent{{\bf Before collision} ($z\rightarrow -\infty$)}\\
\underline{CCS $S_1$}: 
\numparts\bea
\hspace{-1.5cm}q_j^{1-}={2A_j^{1-}} \left(\frac{\mbox{cos}(P_j^{1-})\mbox{cosh}(\eta_{1R}^-)+i~\mbox{sin}(P_j^{1-})\mbox{sinh}(\eta_{1R}^-)}{ 4\mbox{cosh}^2(\eta_{1R}^-)+L^{1-}}\right) e^{i\eta_{1I}}, \quad j=1,2,3,
\eea
where $A_j^{1-}= e^{\frac{\delta_{11}^{(j)}+l_j^--\epsilon_{11}}{2}}$, $P_j^{1-}= {\frac{\delta_{11I}^{(j)}-l_{jI}^-}{2}}$, $l_j^-=\mbox{ln}(\alpha_1^{(j)})$, $L^{1-}=e^{(R_1-\frac{\epsilon_{11}}{2})}-2$, and $\eta_{1R}^-=\eta_{1R}+\frac{\epsilon_{11}}{4}$.\\
\noindent\underline{CCS $S_2$}:
\bea
\hspace{-1.5cm}q_j^{2-}={2A_j^{2-}} \left(\frac{\mbox{cos}(P_j^{2-})\mbox{cosh}(\eta_{2R}^-)+i~\mbox{sin}(P_j^{2-})\mbox{sinh}(\eta_{2R}^-)}{ 4\mbox{cosh}^2(\eta_{2R}^-)+L^{2-}}\right) e^{i\eta_{2I}}, \quad j=1,2,3,
\eea
where $A_j^{2-}= e^{\frac{\phi_1^{(j)}+\mu_{11}^{(j)}-R_4-\epsilon_{11}}{2}}$, $P_j^{2-}= {\frac{\phi_{1I}^{(j)}+\mu_{11I}^{(j)}}{2}}$, $L^{2-}=e^{\theta_{11}-(\frac{R_4+\epsilon_{11}}{2})}-2$, and $\eta_{2R}^-=\eta_{2R}+\frac{R_4-\epsilon_{11}}{4}$.\\
\noindent{{\bf After collision} ($z\rightarrow +\infty$)}\\
\underline{CCS $S_1$}:
\bea
\hspace{-1.5cm}q_j^{1+}={2A_j^{1+}} \left(\frac{\mbox{cos}(P_j^{1+})\mbox{cosh}(\eta_{1R}^+)+i~\mbox{sin}(P_j^{1+})\mbox{sinh}(\eta_{1R}^+)}{4\mbox{cosh}^2(\eta_{1R}^+)+L^{1+}}\right) e^{i\eta_{1I}}, \quad j=1,2,3, \label{3c2ccsa}
\eea
where $A_j^{1+}= \frac{(k_1-k_2)(k_1^*+k_2)}{(k_1^*-k_2^*)(k_1+k_2^*)}A_j^{1-}$, $P_j^{1+}= {\frac{\phi_{2I}^{(j)}-\mu_{22I}^{(j)}}{2}} \equiv P_j^{1-}$, $L^{1+}=e^{(\theta_{22}-\frac{R_4+\epsilon_{22}}{2})}-2 \equiv L^{1-}$, and $\eta_{1R}^+=\eta_{1R}+\frac{R_4-\epsilon_{22}}{4}$.\\
\underline{CCS $S_2$}:
\bea
\hspace{-1.5cm}q_j^{2+}=2A_j^{2+} \left(\frac{\mbox{cos}(P_j^{2+})\mbox{cosh}(\eta_{2R}^+)+i~\mbox{sin}(P_j^{2+})\mbox{sinh}(\eta_{2R}^+)}{ 4\mbox{cosh}^2(\eta_{2R}^+)+L^{2+}}\right) e^{i\eta_{2I}}, \quad j=1,2,3, \label{3c2ccsb}
\eea\endnumparts
where $A_j^{2+}= \frac{(k_1^*-k_2^*)(k_1^*+k_2)}{(k_1-k_2)(k_1+k_2^*)}A_j^{2-}$, $P_j^{2+}= {\frac{\delta_{22I}^{(j)}-\l_{jI}^+}{2}} \equiv P_j^{2-}$, $l_j^+=\mbox{ln}(\alpha_2^{(j)})$,  $L^{2+}=e^{(R_2-\frac{\epsilon_{22}}{2})}-2 \equiv L^{2-}$, and $\eta_{2R}^+=\eta_{2R}+\frac{\epsilon_{22}}{4}$.

\begin{figure}[h]
\centering\includegraphics[width=0.56\linewidth]{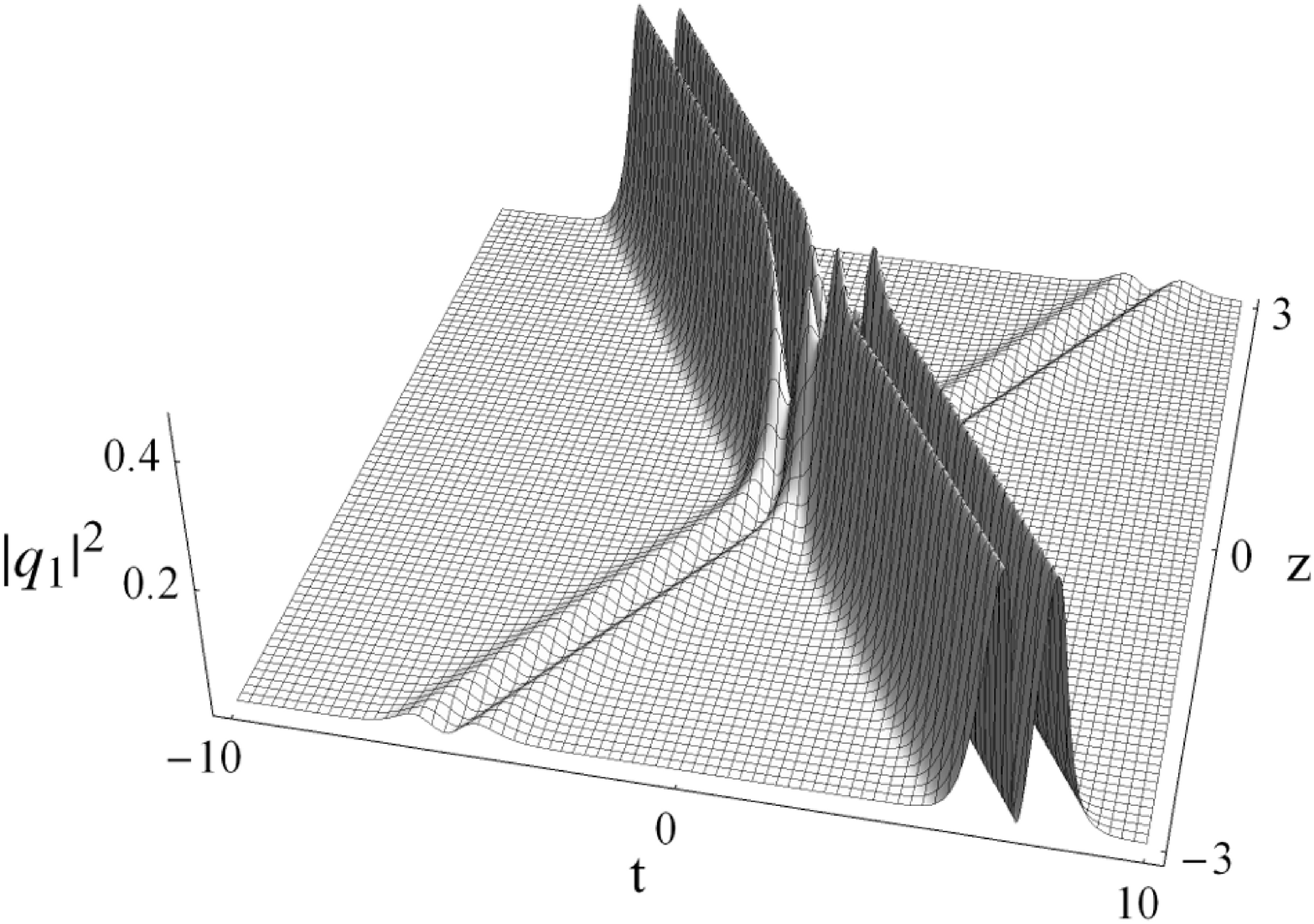}~\\~\includegraphics[width=0.56\linewidth]{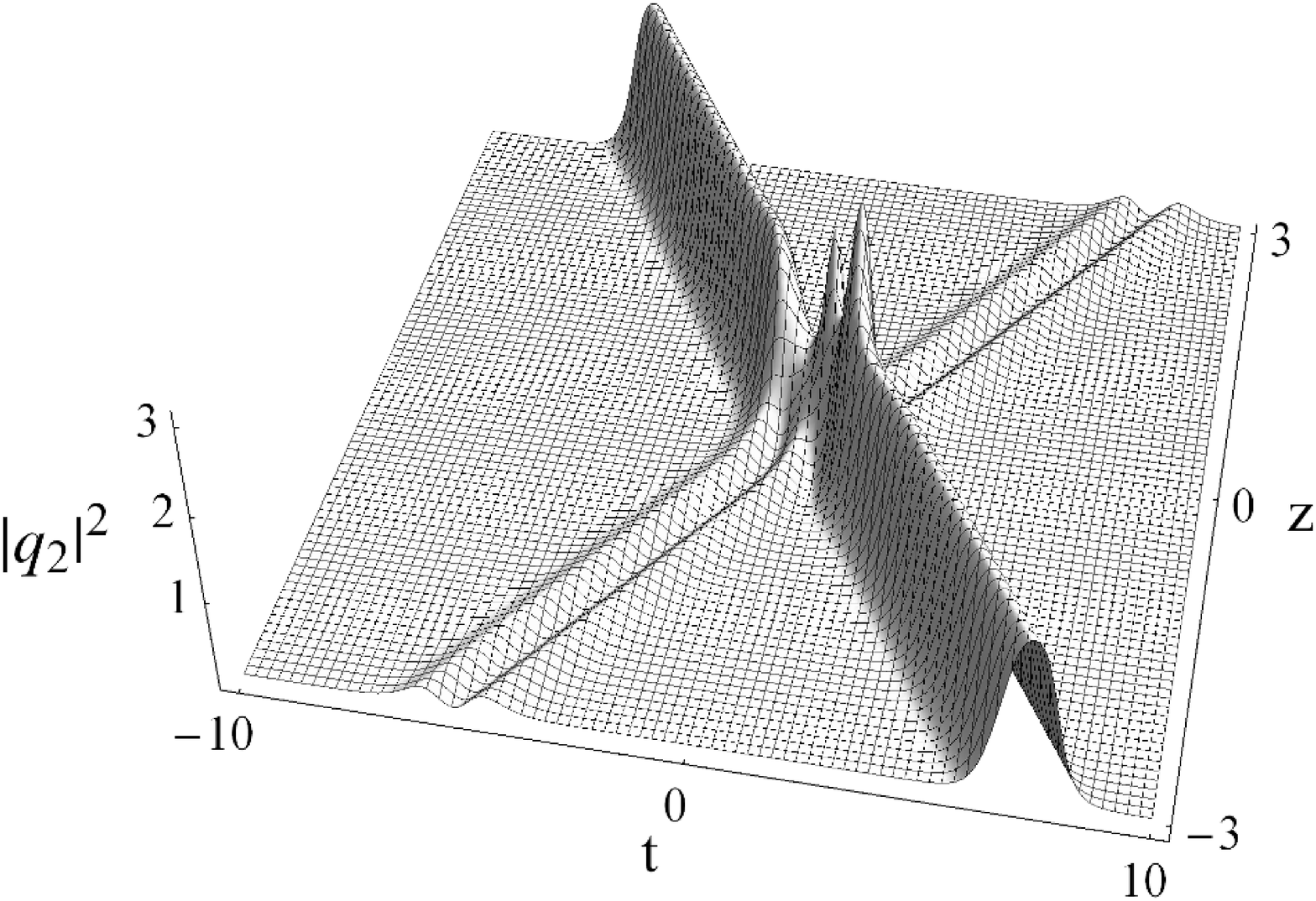}\\~~\includegraphics[width=0.56\linewidth]{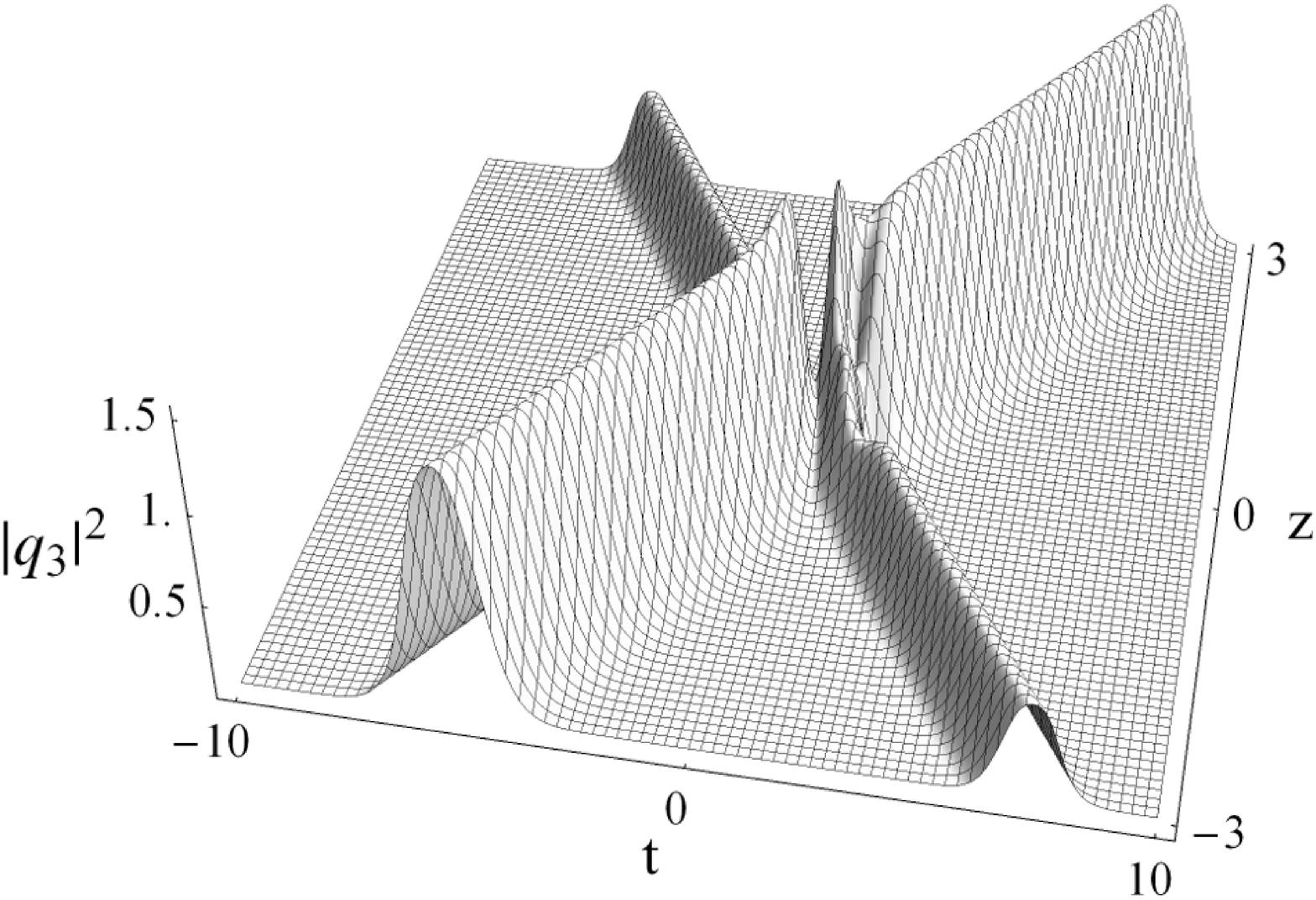}
\caption{Elastic collision of coherently coupled solitons in $3$-CCNLS system.}
\label{3cc2}
\end{figure}

From the expressions for $A_j^{1+}$ and $A_j^{2+}$ given below equations (\ref{3c2ccsa}) and (\ref{3c2ccsb}), respectively, we find that $|A_j^{l+}|^2=|A_j^{l-}|^2,~l=1,2$, and $j=1,2,3$. This displays the elastic nature of collision between two CCSs. Also, the two colliding CCSs $S_1$ and $S_2$ experience phase shifts $\Phi_1=\frac{R_4-\epsilon_{11}-\epsilon_{22}}{4k_{1R}} \equiv \frac{1}{ k_{1R}}\ln \left(\frac{(k_1 - k_2)(k_1^* - k_2^*)}{(k_1^* + k_2)(k_1 + k_2^*)}\right)$ and $\Phi_2=-\left(\frac{k_{1R}}{k_{2R}}\right)\Phi_1$, respectively. In this case, the change in the relative separation distance between the two CCSs is $\Delta t_{12}= \left(1+\frac{k_{1R}}{k_{2R}}\right) \Phi_1$. Here also the phase shifts and hence the relative separation distance of the solitons purely depend on $k$'s only. This kind of elastic collision of the CCSs is shown in figure \ref{3cc2} for $k_1=1.5+i, ~k_2=2-i, ~\gamma=2,~ \alpha_1^{(1)}=0.25,~ \alpha_1^{(2)}=-0.71,~\alpha_1^{(3)}=1.2i,~ \alpha_2^{(1)}=1$, $\alpha_2^{(2)}=1.4i$ and $\alpha_2^{(3)}=0.75i$.

\subsubsection{Collision of incoherently coupled solitons:}
The collision dynamics of incoherently coupled solitons (say $S_1$ and $S_2$) arising for the choice $(\alpha_j^{(1)})^2+(\alpha_j^{(2)})^2+(\alpha_j^{(3)})^2 = 0,~j=1,2,$ discussed here exhibits exciting energy sharing collision behaviour which is not possible in the two-component CCNLS system. The following expressions are the asymptotic forms of the ICSs $S_1$ and $S_2$ during collision.\\
\noindent{\textbf{Before collision} ($z\rightarrow -\infty$)}
\numparts\bea
\hspace{-2.7cm}\left(\begin{array}{c}
  q_j^{1-} \\
  q_j^{2-}
\end{array}\right)^{\bf T} = \frac{1}{2}
    \left(\begin{array}{c}
         A_j^{1-} e^{i\eta_{1I}}\\
        A_j^{2-}e^{i\eta_{2I}}
     \end{array} \right)^{\bf T}
     \left(\begin{array}{cc}
         \mbox{sech}\left(\eta_{1R}+\frac{R_1}{2}\right)~~ & 0 \\
         ~0 & \mbox{sech}\left(\eta_{2R}+\frac{R_3-R_1}{2}\right) \\
       \end{array}\right),
j=1,2,3,~~~~~~~
\eea
where $A_j^{1-}=\alpha_1^{(j)} e^{-\frac{R_1}{2}}$ and $A_j^{2-}=e^{\delta_1^{(j)}-(\frac{R_1+R_3}{2})}$.\\
\noindent{\textbf{After collision} ($z\rightarrow +\infty$)}
\bea
\hspace{-2.7cm}\left(\begin{array}{c}
  q_j^{1+} \\
  q_j^{2+}
\end{array}\right)^{\bf T}= \frac{1}{2}
    \left(\begin{array}{c}
         A_j^{1+} e^{i\eta_{1I}}\\
        A_j^{2+}e^{i\eta_{2I}}
     \end{array} \right)^{\bf T}
     \left(\begin{array}{cc}
         \mbox{sech}\left(\eta_{1R}+\frac{R_3-R_2}{2}\right)~~ & 0 \\
         ~0 & \mbox{sech}\left(\eta_{2R}+\frac{R_2}{2}\right) \\
       \end{array}\right),
j=1,2,3,~~~~~~
\eea\label{3c2ics}\endnumparts
where $A_j^{1+}=e^{\delta_2^{(j)}-(\frac{R_2+R_3}{2})}$ and $A_j^{2+}=\alpha_2^{(j)} e^{-\frac{R_2}{2}}$.  All the other quantities in equation (23) can be obtained from the Appendix for $m=3$.

From the above asymptotic expressions, we arrive at the following expressions relating the amplitudes of ICSs before and after collision.
\numparts\bea
A_j^{1+}=T_j^{(1)}~A_j^{1-}, \quad A_j^{2+}=T_j^{(2)}~A_j^{2-}, \quad j=1,2,3,
\eea
 where the transition amplitudes $T_j^{(1)}$ and $T_j^{(2)}$ of solitons $S_1$ and $S_2$, respectively, are found as
 \bea
T_j^{(1)}&=& \frac{\left(1-\hat\lambda_1 +\frac{\alpha_2^{(j)*} \Omega}{\alpha_1^{(j)}~\kappa_{22}}\right)} {\sqrt{1-\hat\lambda_1~\hat\lambda_2+\frac{|\Omega|^2}{\kappa_{11}\kappa_{22}}}} \left(\frac{(k_1-k_2)(k_1^*+k_2)}{(k_1^*-k_2^*)(k_1+k_2^*)}\right)^{\frac{1}{2}}, ~~ j=1,2,3,\\
T_j^{(2)}&=& - \frac{\sqrt{1-\hat\lambda_1~\hat\lambda_2+\frac{|\Omega|^2}{\kappa_{11}\kappa_{22}}}} {\left(1-\hat\lambda_2+\frac{\alpha_1^{(j)*}\Omega}{\alpha_2^{(j)}~\kappa_{11}}\right)} \left(\frac{(k_1^*-k_2^*)(k_1^*+k_2)}{(k_1-k_2)(k_1+k_2^*)}\right)^{\frac{1}{2}}, ~~ j=1,2,3,
\eea  \label{tra3c}\endnumparts
where $\hat\lambda_1=\frac{\alpha_2^{(j)} \kappa_{12}}{\alpha_1^{(j)} \kappa_{22}}$, $\hat\lambda_2=\frac{\alpha_1^{(j)} \kappa_{21}}{\alpha_2^{(j)} \kappa_{11}} $, and $\Omega=\frac{\gamma }{(k_1-k_2)}\displaystyle\sum_{l=1}^3 (\alpha_1^{(l)}\alpha_2^{(l)})$.

\begin{figure}[h]
\centering\includegraphics[width=0.6\linewidth]{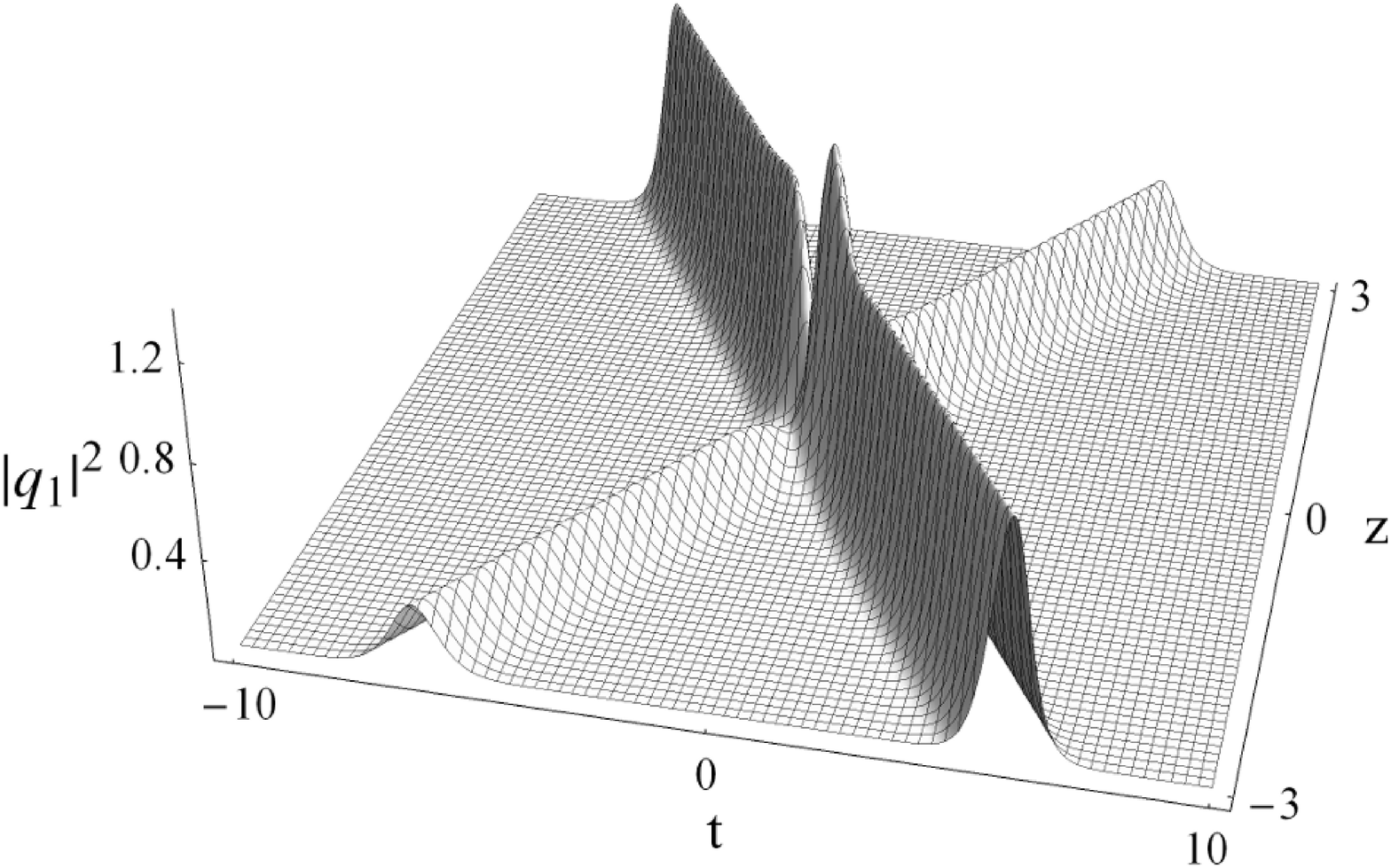}~\\~\includegraphics[width=0.6\linewidth]{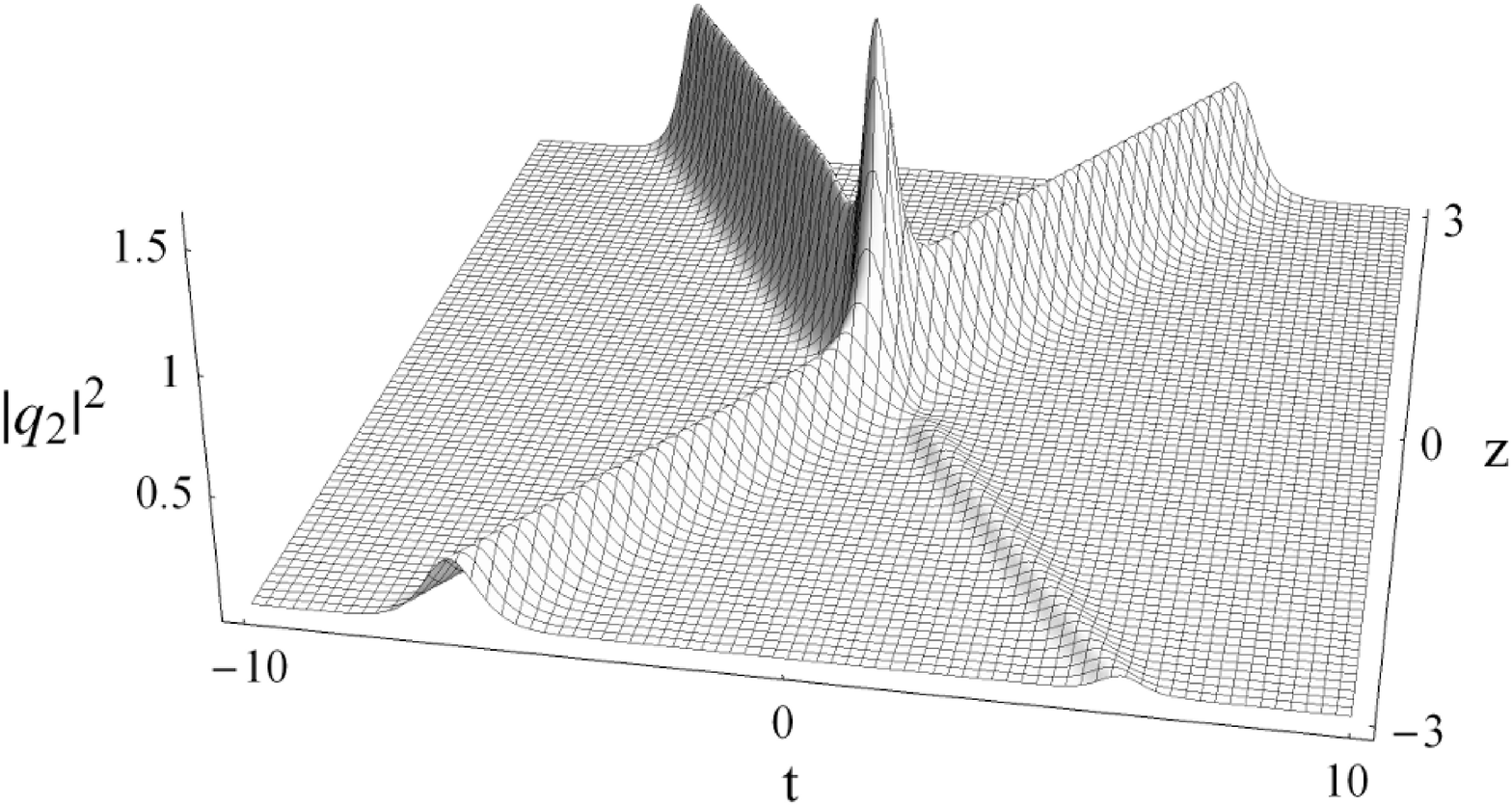}\\~~\includegraphics[width=0.6\linewidth]{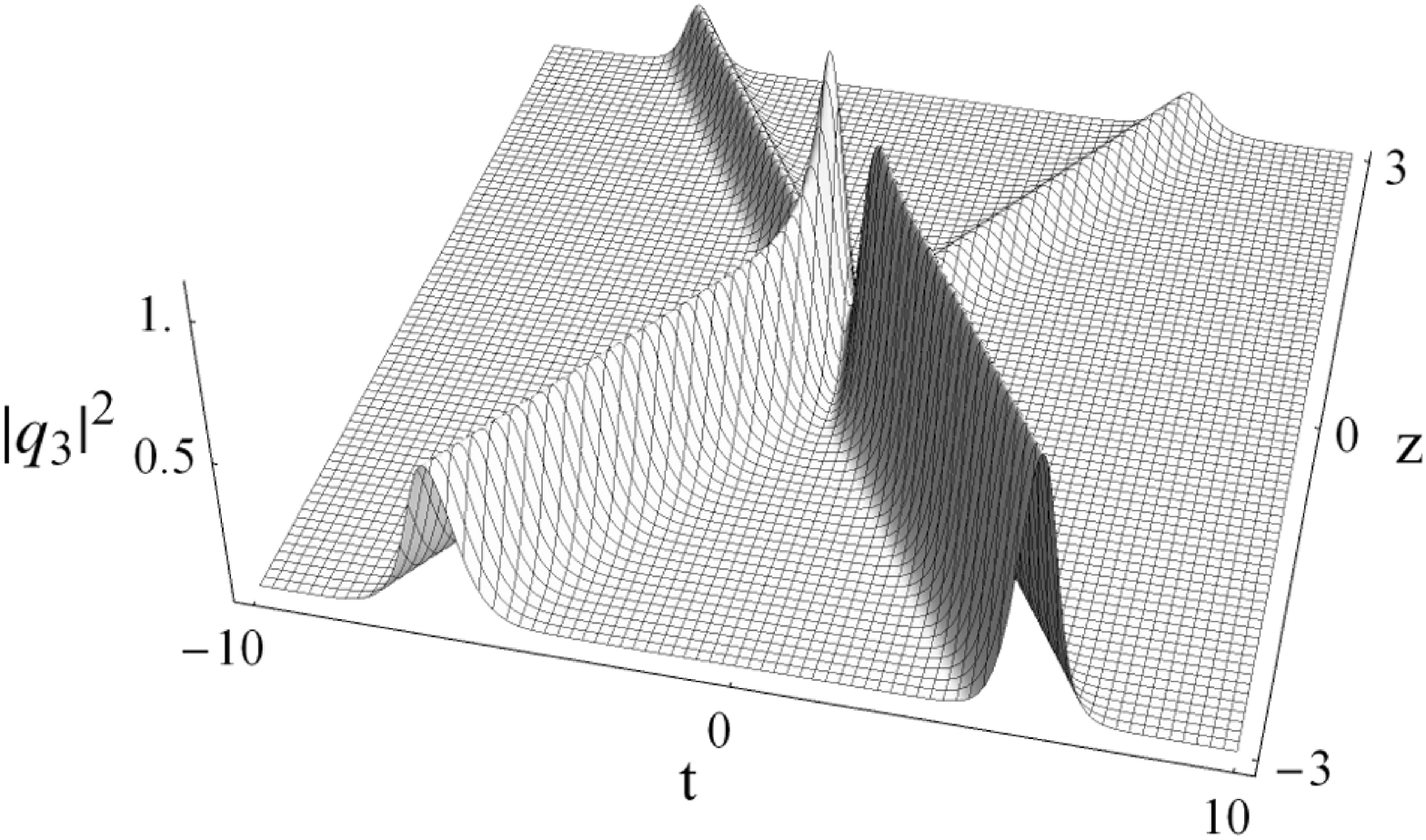}
\caption{Novel shape changing collision of incoherently coupled solitons in $3$-CCNLS system.}
\label{3cc3}
\end{figure}

\begin{figure}[h]
\centering\includegraphics[width=0.6\linewidth]{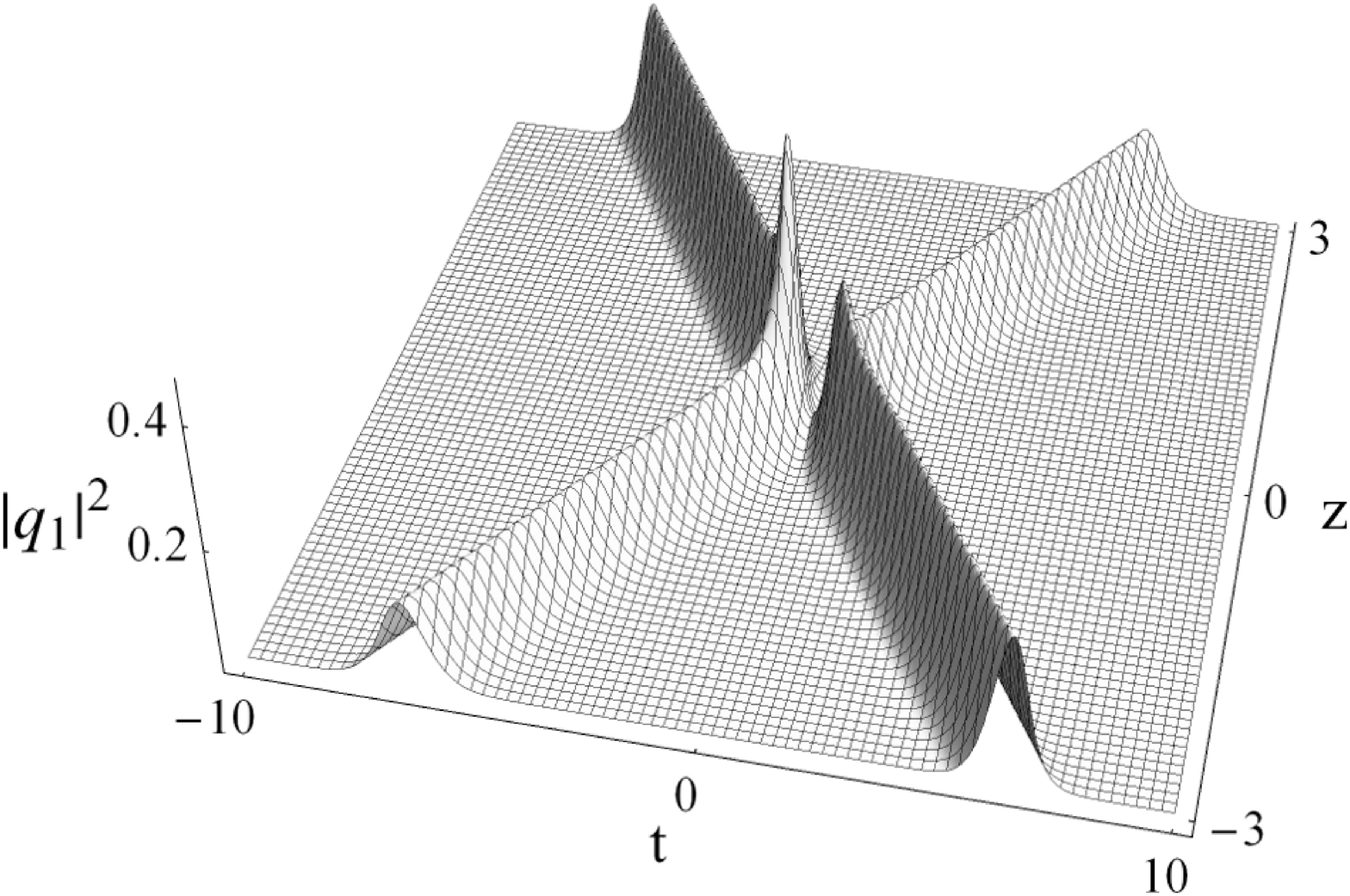}~\\~\includegraphics[width=0.6\linewidth]{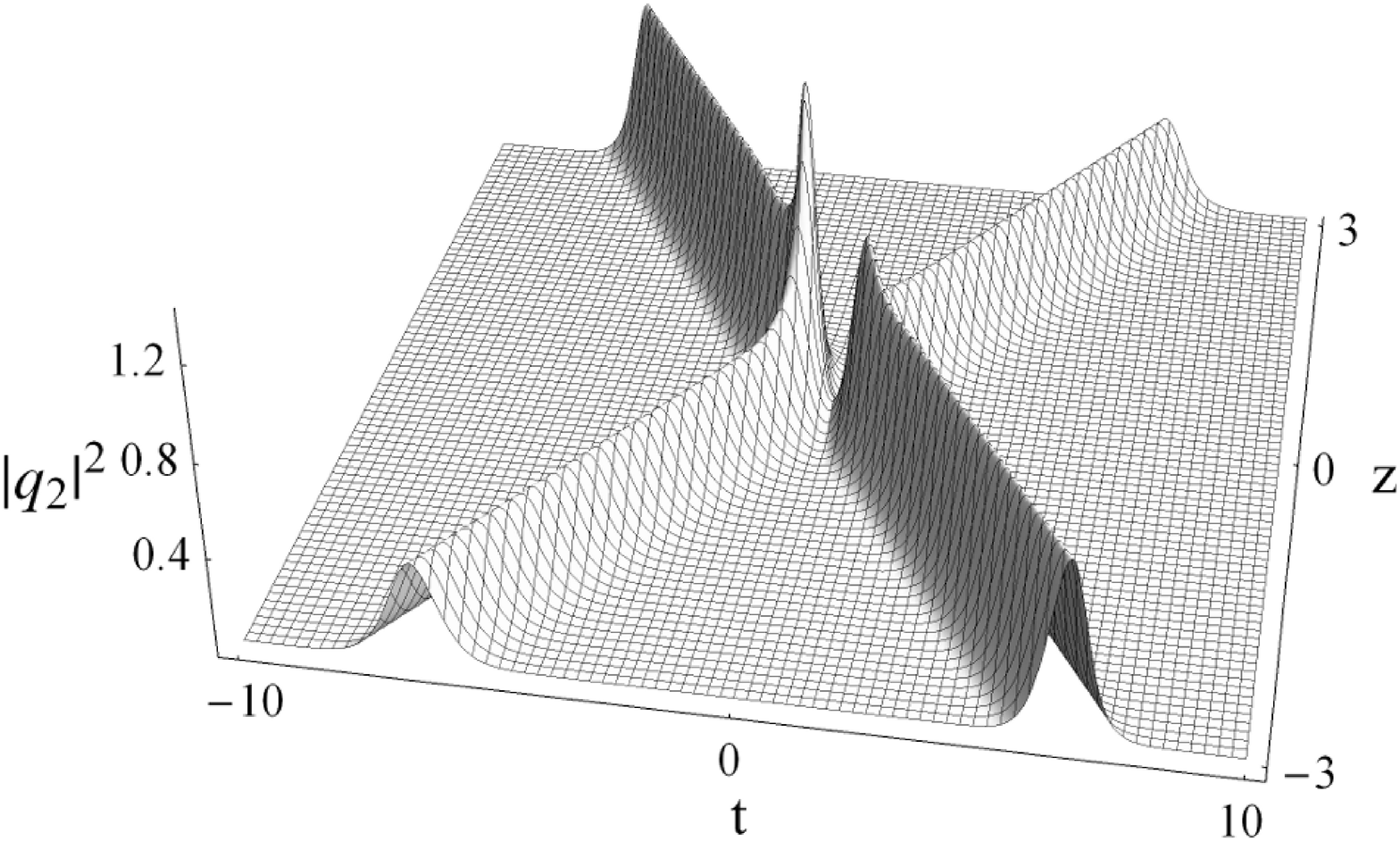}\\~~\includegraphics[width=0.6\linewidth]{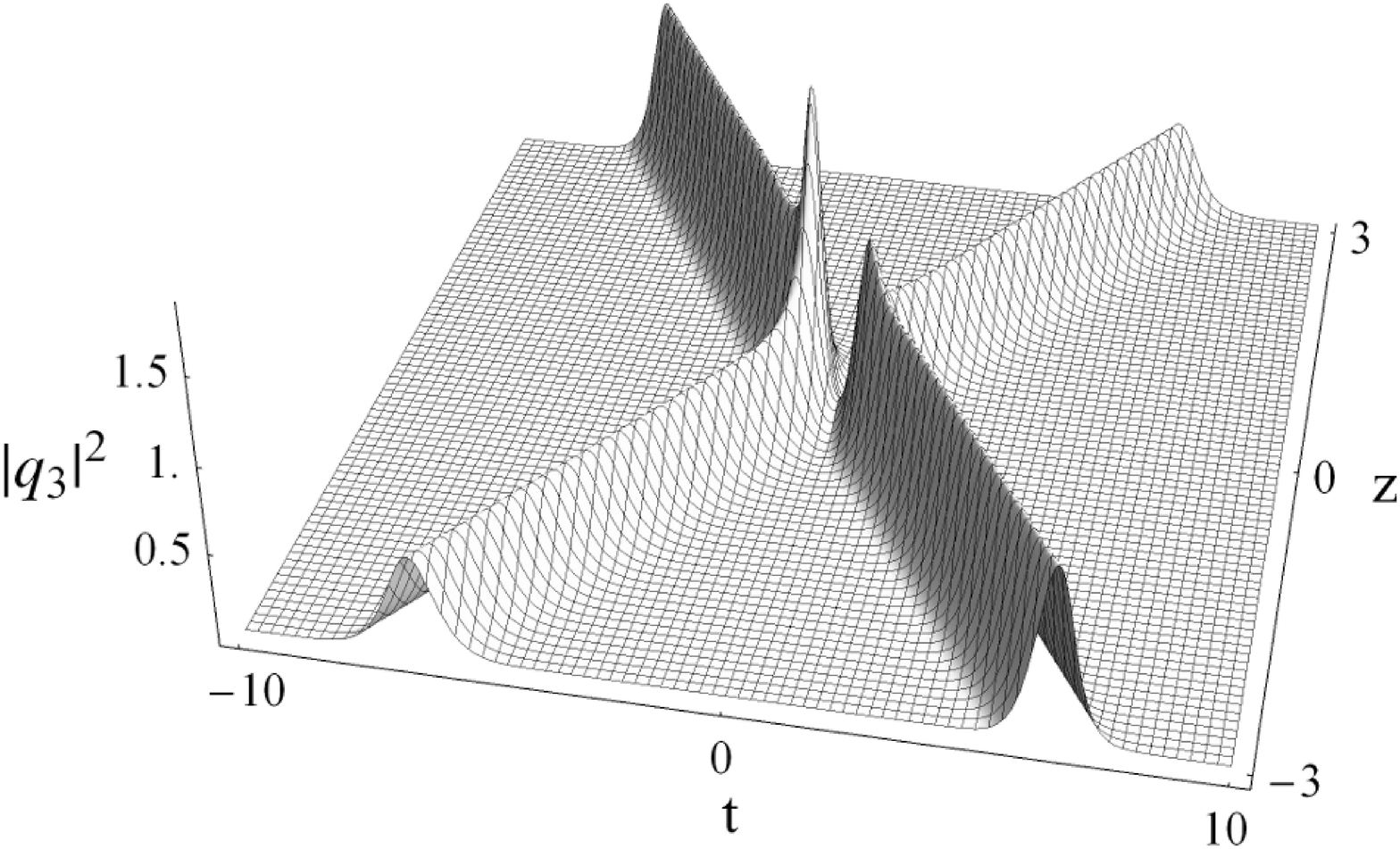}
\caption{Elastic collision of incoherently coupled solitons in $3$-CCNLS system.}
\label{3cc4}
\end{figure}

It can be noticed here that $T_j^{(1)}$ and $T_j^{(2)}$, $j=1,2,3$, are unimodular only for the choice $\frac{\alpha_1^{(1)}}{\alpha_2^{(1)}}=\frac{\alpha_1^{(2)}}{\alpha_2^{(2)}}=\frac{\alpha_1^{(3)}}{\alpha_2^{(3)}}$ and hence the elastic collision occurs only for this choice. Otherwise, the amplitudes of the solitons are different as $|{T_j}^{(l)}|^2\neq 1$, $l=1,2$, $j=1,2,3$, and this results in an amplitude/intensity redistribution among the two solitons split up in the three components. The phase shift experienced by the solitons $S_1$ is $\Phi_1=\frac{R_3-R_2-R_1}{2k_{1R}} \equiv \frac{1}{2k_{1R}}\ln \left[\frac{|k_1-k_2|^2}{|k_1+k_2^*|^2} \left(1-\hat\lambda_1~\hat\lambda_2+\frac{|\Omega|^2}{\kappa_{11}\kappa_{22}}\right)\right]$ and that of $S_2$ is found to be $\Phi_2=-\frac{k_{1R}}{k_{2R}}\Phi_1$. Here the change in the relative separation distance between the two solitons become $\Delta t_{12}= \left(1+\frac{k_{1R}}{k_{2R}}\right) \Phi_1$. In contrary to the collision scenario in the two component case and also to the two other collision processes discussed in this subsection 6.2, here the phase-shift does depend on $\alpha$-parameters also in addition to $k$'s. Hence the phase shift and the change in the relative separation distance between the colliding solitons can be tuned by altering the polarization parameter $\alpha$ and $k$ suitably. This collision scenario is similar to the shape changing collision of solitons involving energy sharing among the colliding solitons in the three-component Manakov system reported in refs. \cite{{tkprl},{rk}} and will have important applications in the context of optical computing and also in achieving multi-state logic \cite{{tkpre},{abloinv}}. Figure \ref{3cc3} shows a typical shape changing (energy sharing) collision in which the intensity of solitons $S_1$ and $S_2$ is enhanced in $q_1$ and $q_2$ components and suppressed in $q_3$ component after collision. Standard elastic collision is depicted in figure \ref{3cc4}. The parameters for figure \ref{3cc3} are $k_1=1.5+i, ~k_2=2-i, ~\gamma=2,~ \alpha_1^{(1)}=\sqrt{2},~ \alpha_1^{(2)}=\sqrt{2},~\alpha_1^{(3)}=2i,~ \alpha_2^{(1)}=\sqrt{8}~i$, $\alpha_2^{(2)}=\sqrt{6}$ and $\alpha_2^{(3)}=\sqrt{2}$. In figure\ref{3cc4} $k_1$, $k_2$, $\gamma$ are chosen as in figure\ref{3cc3} and $\alpha_{1,2}^{(1)}=\sqrt{2},~ \alpha_{1,2}^{(2)}=\sqrt{6},~\alpha_{1,2}^{(3)}=\sqrt{8}~i$.

This kind of shape changing (energy sharing) collision between the incoherent solitons is not at all possible in the two-component CCNLS system as the constraint $(\alpha_j^{(1)})^2+(\alpha_j^{(2)})^2=0$, $j=1,2$, restricts them to behave like standard NLS solitons. But when we go for three-component the additional freedom involved due to the presence of the third component allows the solitons to behave like three-component Manakov solitons and results in fascinating energy sharing collisions accompanied by phase-shifts which can be tuned by altering both $k$'s and $\alpha$-parameters. The shape changing collision between coherent soliton and an incoherent soliton in three-component CCNLS is also different from the above discussed collision scenario in the sense that in the former case energy switching occurs only in a particular soliton (CCS) and the energy of individual components is not conserved whereas the total energy is conserved. Also it is not possible to achieve elastic collisions during the collision between CCS and ICS. But in the present case, energy redistribution occurs in both solitons and also the energy of individual components as well as the total energy are conserved.

\subsection{\textbf{Soliton collisions of $m$-component CCNLS equations}}
The above analysis of 2-component and 3-component CCNLS system can be extended to arbitrary $m$-component CCNLS system. The asymptotic expressions for the three different kinds of soliton collisions for the $m$-component case, with arbitrary $m$, can be obtained from equations (21)--(23) by allowing $j$ and $l$ to run from $1$ to $m$ and from the expressions given in the Appendix. The collision between the $m$-component coherently coupled solitons again turns out to be standard elastic collision with amplitude dependent phase shift and change in relative separation distance. But the $m$-component CCS exhibits intensity switching whenever it undergoes collision with $m$-component ICS and experiences amplitude dependent phase shift which in turn results in change in the relative separation distance between the solitons. The transition amplitude for the $m$-component CCS is given by
\numparts \bea
\hspace{-2cm}T_j=\left(\frac{(k_1^*+k_2)(k_1-k_2)\big|(\alpha_1^{(j)} \kappa_{22}-\alpha_2^{(j)} \kappa_{12})+\alpha_2^{(j)*} \Omega\big|^2}{(k_1+k_2^*) (k_1^*-k_2^*) ~\kappa_{22}^2 ~ |\alpha_1^{(j)}|^2}\right)^{\frac{1}{2}}, ~~ j=1,2,3,...,m,
\eea
where $\Omega=\frac{\gamma }{(k_1-k_2)}\displaystyle\sum_{l=1}^m (\alpha_1^{(l)}\alpha_2^{(l)})$. The expressions for phase shift and the change in relative separation distance are found to be $\Phi_1=\frac{\theta_{11}-R_2-\epsilon_{11}}{4k_{1R}} \equiv \frac{1}{ k_{1R}}\ln \left(\frac{(k_1 - k_2)(k_1^* - k_2^*)}{(k_1^* + k_2)(k_1 + k_2^*)}\right)$ and $\Delta t_{12}= \left(1+\frac{2 k_{1R}}{k_{2R}}\right) \Phi_1$, respectively. But the incoherently coupled soliton exhibits elastic collision in all the $m$ components only with amplitude dependent phase shift.

An interesting energy sharing collision takes place between the $m$-component incoherently coupled solitons with $m>2$ which is not possible in their two-component counterparts. In this case both the ICSs undergo shape changing collision characterized by intensity redistribution and an amplitude dependent phase shift in all the components, which is similar to the soliton collisions of multicomponent Manakov solitons \cite{tkprl}. The corresponding transition amplitudes relating the amplitudes of the solitons before and after collision can be expressed as
\bea
\hspace{-2cm}T_j^{(1)}&=& \frac{\left(1-\hat\lambda_1 +\frac{\alpha_2^{(j)*} \Omega}{\alpha_1^{(j)}~\kappa_{22}}\right)} {\sqrt{1-\hat\lambda_1~\hat\lambda_2+\frac{|\Omega|^2}{\kappa_{11}\kappa_{22}}}} \left(\frac{(k_1-k_2)(k_1^*+k_2)}{(k_1^*-k_2^*)(k_1+k_2^*)}\right)^{\frac{1}{2}}, ~~ j=1,2,3,...,m,\\
\hspace{-2cm}T_j^{(2)}&=& - \frac{\sqrt{1-\hat\lambda_1~\hat\lambda_2+\frac{|\Omega|^2}{\kappa_{11}\kappa_{22}}}} {\left(1-\hat\lambda_2+\frac{\alpha_1^{(j)*}\Omega}{\alpha_2^{(j)}~\kappa_{11}}\right)} \left(\frac{(k_1^*-k_2^*)(k_1^*+k_2)}{(k_1-k_2)(k_1+k_2^*)}\right)^{\frac{1}{2}}, ~~ j=1,2,3...,m,
\eea  \endnumparts
where $\hat\lambda_1=\frac{\alpha_2^{(j)} \kappa_{12}}{\alpha_1^{(j)} \kappa_{22}}$, $\hat\lambda_2=\frac{\alpha_1^{(j)} \kappa_{21}}{\alpha_2^{(j)} \kappa_{11}} $, and $\Omega=\frac{\gamma }{(k_1-k_2)}\displaystyle\sum_{l=1}^m (\alpha_1^{(l)}\alpha_2^{(l)})$. Also, the amplitude dependent phase shift of the solitons $S_1$ and $S_2$ can be written as $\Phi_1=\frac{R_3-R_2-R_1}{2k_{1R}} \equiv \frac{1}{2k_{1R}}\ln \left[\frac{|k_1-k_2|^2}{|k_1+k_2^*|^2} \left(1-\hat\lambda_1~\hat\lambda_2+\frac{|\Omega|^2}{\kappa_{11}\kappa_{22}}\right)\right]$ and $\Phi_2=-\frac{k_{1R}}{k_{2R}}\Phi_1$, respectively. The above transition amplitudes become unimodular and result in elastic collision  when the $\alpha$-parameters satisfy the specific condition $\frac{\alpha_1^{(1)}}{\alpha_2^{(1)}}=\frac{\alpha_1^{(2)}}{\alpha_2^{(2)}}=\cdots=\frac{\alpha_1^{(m)}}{\alpha_2^{(m)}}$.

\section{Conclusion}
In this paper we have considered the integrable $m$-component CCNLS system (\ref{e2nc}) describing simultaneous propagation of $m$ fields in Kerr type nonlinear media and obtained the correct bilinear equations by a non-standard bilinearization procedure, resulting in more general soliton solutions. The solitons are classified as coherently coupled solitons and incoherently coupled solitons depending upon the presence and absence of coherent contribution from the co-propagating components/modes. We show that one can get coherently coupled or incoherently coupled solitons by tuning the polarization parameters suitably. Apart from this, one can also adopt the non-standard Hirota's bilinearization method to construct exact soliton solutions of similar kind of multicomponent systems which arising in the context of nonlinear optics, spinor condensates, etc. Then by considering more general two-soliton solutions we have analyzed the collisions among them. Our analysis on their collision dynamics explores several interesting collision properties. For the two-component case, we find that the CCS(ICS) undergoes elastic collision with CCS (ICS). But during its collision with ICS, the CCS experiences energy switching along with amplitude dependent phase shift and change in relative separation distance depending on $k$-parameters, leaving the ICS unaffected. Our study on the soliton collisions in the three-component CCNLS system reveals the fact that the collision dynamics is similar to the two-component case when one considers the collision between two CCSs or collision of CCS with ICS. But the collision between two ICSs in $3$-component CCNLS system displays entirely different behaviour from that of two-component CCNLS system. Here both the colliding solitons experience energy redistribution and an amplitude dependent phase shift along with change in the relative separation distance which depends on both $k$'s as well as $\alpha$-parameters. This kind of collision scenario is similar to the three-component Manakov soliton collisions. Our analysis on the $m$-component case, with arbitrary $m$, also shows that the collision between CCSs is elastic and always there occurs energy switching in the CCS during its collision with ICS. We have also pointed out that the collision scenario between ICSs involving energy sharing among the solitons in all components for $m>2$ and is completely different from $m=2$ case, where the collision is mere elastic. However the collision between ICSs for $m>2$ case, can be made elastic for specific choice of $\alpha$-parameters. This study will find applications in soliton collision based optical computing and in optical switches. We believe that our results will also have important ramifications in nonlinear optics and in multicomponent Bose-Einstein condensates.

\ack
The authors gratefully acknowledge the support of the Department of Science and Technology,
Government of India under a major research project. T.K. and K.S. thank Professor M.Lakshmanan and M.Vijayajayanthi for useful discussions. The authors also thank the principal and management of Bishop Heber College for constant support and encouragement.
\section*{Appendix}
The various quantities appearing in the sections 3--6 are defined below. 
\bea
e^{R_u}&=&\frac{\kappa _{uu}}{(k_u+k_u^*)},~~ e^{\delta _0}=\frac{ \kappa _{12}}{(k_1+k_2^*)},~~
 e^{\delta _0^*}=\frac{ \kappa _{21}}{(k_2+k_1^*)},~~\nonumber\\
e^{\delta _{uv}^{(j)}}&=&\frac{\gamma \alpha_v^{(j)*}\displaystyle\sum_{l=1}^m (\alpha_u^{(l)})^2}{2 (k_u+k_v^*)^2},~~e^{\delta _u^{(j)}}=\frac{\gamma \alpha _u^{(j)*} \displaystyle\sum_{l=1}^m (\alpha _1^{(l)} \alpha _2^{(l)})+(k_1-k_2) (\alpha _1^{(j)} \kappa _{2u}-\alpha _2^{(j)} \kappa_{1u})}{(k_1+k_u^*) (k_2+k_u^*)},\nonumber\\
e^{\epsilon _{uv}}&=&\frac{ \gamma ^2 \displaystyle\sum_{j=1}^m(\alpha _u^{(j)})^2 \displaystyle\sum_{j=1}^m(\alpha _v^{(j)*})^2}{4 (k_u+k_v^*)^4},~\quad e^{\tau _u}=\frac{\gamma ^2 \displaystyle\sum_{j=1}^m(\alpha _1^{(j)*} \alpha _2^{(j)*}) \displaystyle\sum_{j=1}^m(\alpha _u^{(j)})^2}{2 (k_u+k_1^*)^2 (k_u+k_2^*)^2},\nonumber\\
e^{\lambda _{uv}}&=&\frac{(k_1-k_2)^2 \kappa_{uv} \displaystyle\sum_{j=1}^m(\alpha _{3-u}^{(j)})^2}{(k_u+k_v^*)(k_{3-u}+k_v^*)^2},~\nonumber\\
e^{\mu _{uv}^{(j)}}&=&\frac{\gamma^2 (k_1-k_2)^2 \alpha _{3-u}^{(j)} \displaystyle\sum_{l=1}^m(\alpha _u^{(l)})^2 \displaystyle\sum_{l=1}^m(\alpha _v^{(l)*})^2}{4 (k_u+k_v^*)^4 (k_{3-u}+k_v^*)^2},\nonumber\\
e^{\mu_u^{(j)}}&=&\frac{(k_1-k_2)^2 \gamma ^2}{2\tilde{D}}{\displaystyle\sum_{l=1}^m (\alpha_u^{(l)})^2} \left(\left[(k_{3-u}+k_1^*)^2+(k_2^*-k_1^*)(k_{3-u}+k_2^*)\right] \alpha _{3-u}^{(j)} \alpha _1^{(j)*} \alpha _2^{(j)*}\right.\nonumber\\
&& \left.+(k_{3-u}+k_1^*) (k_1^*-k_2^*) \alpha _1^{(j)*}\displaystyle\sum_{{l=1,l\neq j}}^m (\alpha_{3-u}^{(l)} \alpha_2^{(l)*}) \right.\nonumber\\
&& \left.-(k_1^*-k_2^*) (k_{3-u}+k_2^*) \alpha _2^{(j)*} \displaystyle\sum_{{l=1,\l\neq j}}^m (\alpha_{3-u}^{(l)} \alpha_1^{(l)*}) \right.\nonumber\\
&& \left.+ (k_{3-u}+k_1^*) (k_{3-u}+k_2^*) \alpha_{3-u}^{(j)} \displaystyle\sum_{{l=1,l\neq j}}^m (\alpha_1^{(l)*} \alpha_2^{(l)*}) \right),\nonumber\\
e^{\theta _{uv}}&=&\frac{\gamma ^3(k_1-k_2)^2 (k_1^*-k_2^*)^2}{{4\tilde{D}}(k_u+k_v^*)^2} \displaystyle\sum_{j=1}^m(\alpha _u^{(j)})^2 \displaystyle\sum_{j=1}^m(\alpha _v^{(j)*})^2 \displaystyle\sum_{j=1}^m(\alpha _{3-u}^{(j)} \alpha _{3-v}^{(j)*}),\nonumber\\
e^{\lambda _u}&=&\frac{\gamma ^2(k_1-k_2)^4 \displaystyle\sum_{j=1}^m(\alpha_1^{(j)})^2 \displaystyle\sum_{j=1}^m(\alpha_2^{(j)})^2 \displaystyle\sum_{j=1}^m(\alpha_u^{(j)*})^2}{4 (k_1+k_u^*)^4 (k_2+k_u^*)^4},\nonumber\\
e^{\lambda _3}&=&\frac{\gamma^2(k_1-k_2)^4}{2\tilde{D}} \displaystyle\sum_{j=1}^m(\alpha_1^{(j)})^2 \displaystyle\sum_{j=1}^m(\alpha _2^{(j)})^2 \displaystyle\sum_{j=1}^m(\alpha _1^{(j)*} \alpha_2^{(j)*}), \nonumber\\
e^{\phi_u^{(j)}}&=&\frac{\gamma ^3 (k_1-k_2)^4 (k_1^*-k_2^*)^2} {8\tilde{D} {(k_1+k_u^*)^2(k_2+k_u^*)^2}} {\alpha _{3-u}^{(j)*}\displaystyle\sum_{l=1}^m(\alpha _1^{(l)})^2 \displaystyle\sum_{l=1}^m(\alpha _2^{(l)})^2 \displaystyle\sum_{l=1}^m(\alpha_u^{(l)*})^2},\nonumber\\
e^{R_3}&=& \frac{|k_1-k_2|^2(\kappa_{11}\kappa_{22}-\kappa_{12}\kappa_{21})+\gamma^2\Big| \displaystyle\sum_{j=1}^m (\alpha_1^{(j)}\alpha_2^{(j)})\Big|^2}{(k_1+k_1^*)|k_1+k_2^*|^2(k_2+k_2^*)}, \nonumber\\
e^{R_4}&=&\frac{\gamma ^4 |k_1-k_2|^8}{16\tilde{D}^2} \displaystyle\sum_{j=1}^m (\alpha _1^{(j)})^2 \displaystyle\sum_{j=1}^m (\alpha _1^{(j)*})^2 \displaystyle\sum_{j=1}^m (\alpha _2^{(j)})^2 \displaystyle\sum_{j=1}^m (\alpha _2^{(j)*})^2,~~~~~\nonumber
\eea
where
\bea
\tilde{D}&=&(k_1+k_1^*)^2(k_1^*+k_2)^2 (k_1+k_2^*)^2 (k_2+k_2^*)^2, \nonumber\\
\kappa_{uv}&=&\frac{\gamma \displaystyle\sum_{j=1}^m(\alpha_u^{(j)} \alpha_v^{(j)*})}{(k_u+k_v^*)}.\nonumber
\eea
Here $u,v=1,2$ and $j,l=1,2,3,...,m$.

\end{document}